\documentclass[epj]{svjour}
\usepackage{amsmath,amssymb,amsfonts}	
\usepackage{graphicx}
\usepackage{xspace}
\usepackage{booktabs}
\usepackage{slashed}
\usepackage{makecell}
\usepackage[switch]{lineno} 
\usepackage[colorlinks=true,pdfstartview=FitV, linkcolor=blue, citecolor=blue,urlcolor=blue]{hyperref} 
\usepackage[sort&compress,numbers,colon,merge]{natbib}

\def\PK {\ensuremath{\mathrm{K}}\xspace}
\def\kaon {{\ensuremath{\PK}}\xspace}
\def\KS {{\ensuremath{\kaon^0_{\mathrm{S}}}}\xspace}

\def\Lz {{\ensuremath{\mathrm{\Lambda}}}\xspace}

\begin{document}

%\linenumbers
%
\title{LHCb potential to discover long-lived new physics particles with lifetimes above 100\,ps}
\titlerunning{LHCb potential to discover new LLPs with lifetimes above 100 ps}
\authorrunning{V.~Gorkavenko \textit{et al}.}
\author{Volodymyr Gorkavenko\inst{1}\thanks{\emph{gorkavol@gmail.com}}, 
Brij Kishor Jashal\inst{2,3}\thanks{\emph{brij.kishor.jashal@cern.ch}}, 
Valerii Kholoimov\inst{1,2}\thanks{\emph{valerii.kholoimov@cern.ch}},
Yehor Kyselov\inst{1}\thanks{\emph{kiselev883@gmail.com}},
Diego Mendoza\inst{2}\thanks{\emph{dmendoza@cern.ch}},
Maksym Ovchynnikov\inst{4}\thanks{\emph{maksym.ovchynnikov@cern.ch}},
Arantza Oyanguren\inst{2}\thanks{\emph{Arantza.Oyanguren@ific.uv.es}}, 
Volodymyr Svintozelskyi\inst{1,2}\thanks{\emph{volodymyr.svintozelskyi@cern.ch}}, 
Jiahui Zhuo\inst{2}\thanks{\emph{Jiahui.Zhuo@ific.uv.es}}
}                     

\institute{Taras Shevchenko National University of Kyiv, Kyiv, Ukraine 
\and IFIC, Universitat de Val$\grave{e}$ncia-CSIC, Apt. Correus 22085, E-46071 Val$\grave{e}$ncia, Spain 
\and TIFR, Tata Institute of Fundamental Research, Mumbai, India 
\and KIT, Institut für Astroteilchen Physik, Karlsruher Institut für Technologie, Germany}

\abstract{For years, it has been believed that the main LHC detectors can only restrictively play the role of a lifetime frontier experiment exploring the parameter space of long-lived particles (LLPs) -- hypothetical particles with tiny couplings to the Standard Model. This paper demonstrates that the LHCb experiment may become a powerful lifetime frontier experiment if it uses the new \texttt{Downstream} algorithm reconstructing tracks that do not let hits in the LHCb vertex tracker. In particular, for many LLP scenarios, LHCb may be as sensitive as the proposed experiments beyond main LHC detectors for various LLP models, including heavy neutral leptons, dark scalars, dark photons, and axion-like particles.}

\PACS{{}{IMSc/2023/06/09}}   

\maketitle

\section{Introduction}
\label{sec:intro}
The Standard Model (SM) of particle physics stands as a robust and well-established theory, providing a framework for understanding the fundamental particles and their interactions. Despite its impressive success over more than five decades, the SM falls short in explaining numerous observed phenomena across the realms of particle physics, astrophysics, and cosmology.
One avenue of extending the SM involves the introduction of particles with masses below the electroweak scale that interact with SM particles. These interactions are mediated by operators referred to as ``portals''~\cite{Alekhin:2015byh}. Accelerator experiments have already ruled out large coupling strengths for such particles, earning them the moniker ``Feebly Interacting Particles''. Small coupling means long lifetimes, and therefore, they are also referred to as long-lived particles (LLPs). The concept of LLPs has gained increasing prominence in the last decade, as evidenced by a growing body of literature (see~\cite{Alekhin:2015byh, Beacham:2019nyx, Antel:2023hkf} and related references), with numerous experimental efforts dedicated to their discovery.

Initially, the primary approach to investigating LLPs involved utilizing the LHC's main detectors, namely CMS, ATLAS, and LHCb. However, these ongoing searches at the LHC face notable limitations that hinder their efficacy in probing LLPs~\cite{CMS:2022fut,ATLAS:2022atq,LHCb:impact}. For instance, the inner trackers have relatively small dimensions, restricting the effective decay volume and, consequently, the probability of LLP decays occurring within it. Additionally, the proximity of these trackers to the production point results in substantial background contamination, necessitating stringent selection criteria that inevitably reduce the number of detectable LLP-related events. Another challenge arises from the limitations imposed by current triggering mechanisms, which require tagging of events at the LLP production vertex, often necessitating the presence of a high-$p_{T}$ lepton, meson, or associated jets. This pre-selection process further curtails the event rate with LLPs and constrains the range of LLP models amenable to investigation. For instance, the main production mode for GeV-scale Heavy Neutral Leptons ($N$) involves the decay $B\to \ell+N$, where the momentum of the lepton $\ell$ is insufficient for triggering.

Recognizing these constraints, the scientific community has begun exploring alternative experiments beyond the confines of the LHC detectors~\cite{Beacham:2019nyx}, encompassing both collider-based setups situated near the LHC and beam dump experiments adopting a displaced decay volume concept. These latter experiments employ an extracted beam line aimed at a stationary target, offering greater flexibility in terms of geometric dimensions and circumventing the limitations imposed by the existing LHC detector searches.

Furthermore, in response to the challenges of detecting LLP’s various innovative ideas have emerged to enhance the capabilities of ATLAS, CMS, and LHCb searches~\cite{Shchutska:2023cgi}. These proposals encompass track-triggers that obviate the need for production vertex tagging and exploit displaced sections of the detector as an effective decay volume. For example, Ref.~\cite{CMS-PAS-EXO-22-017} explores the possibility of detecting decays occurring within the CMS muon chamber, albeit still requiring the presence of a high-$p_{T}$ prompt lepton.

This paper presents a method to significantly augment the reach of the LHCb experiment for probing LLPs by harnessing novel algorithms developed under the new LHCb trigger software scheme~\cite{CERN-LHCC-2020-006}. In particular, the newly introduced \texttt{Downstream} algorithm \cite{mrdownstream} emerges as a pivotal tool for extending the search for LLPs with decay lifetimes significantly exceeding 100 ps.

The paper's structure is as follows: In Sec.\ref{sec:lhcb}, we delve into the LHCb experiment, the trigger system, and the novel \texttt{Downstream} algorithm. Sec.~\ref{sec:signal_char} outlines expected signal signatures, encompassing production and decay modes specific to various models, while discussing the LHCb experiment's capacity to detect them. Sec.~\ref{sec:backgrounds} scrutinizes anticipated background sources that could influence the search for Beyond the Standard Model (BSM) particles. Sec.~\ref{sec:signal} provides an estimate of the signal yield, including a breakdown of anticipated efficiencies, along with a qualitative comparison with other experimental proposals. Sec.~\ref{sec:sensitivity} presents the sensitivities of the LHCb experiment, incorporating the \texttt{Downstream} algorithm across various LLP scenarios. Finally, Sec.~\ref{sec:conclusions} concludes the paper.

\section{The LHCb experiment}
\label{sec:lhcb}
The LHCb forward spectrometer is one of the main detectors at the Large Hadron Collider (LHC) accelerator at CERN, with the primary purpose of searching for new physics through studies of CP-violation and heavy-flavour hadron decays. It has been operating during its Run~1 (2011-2012) and Run~2 (2015-2018) periods with very high performance, recording an integrated luminosity of 9 $\text{fb}^{-1}$ at center-of-mass energies of 7, 8, and 13 TeV and delivering a plethora of accurate physics results and new particles discoveries.

The upgraded LHCb detector, operational at present during the Run~3 of the LHC, has implied a major change in the experiment. The detectors have been almost completely renewed to allow running at an instantaneous luminosity five times larger than that of the previous running periods, in particular using new readout architectures. A full software trigger executed on Graphic Processor Units (GPU) also represents one of the main features of the new LHCb design, allowing the reconstruction and selection of events in real-time and widening the physics reach of the experiment. The main characteristics of the new LHCb detector are detailed in \cite{LHCbU1}, and summarised in the following. As compared to the previous detector \cite{LHCb_2008}, one of the most important improvements concerns the new tracking system. The LHCb is comprised of a three subdetector tracking system (VErtex LOcator, Upstream Tracker, and SciFi tracker), a particle identification system, based on two-ring imaging Cherenkov detectors, hadronic and electromagnetic calorimeters, and four muon chambers. 

\begin{figure}[h!]
    \centering
    \includegraphics[width=0.5\textwidth]{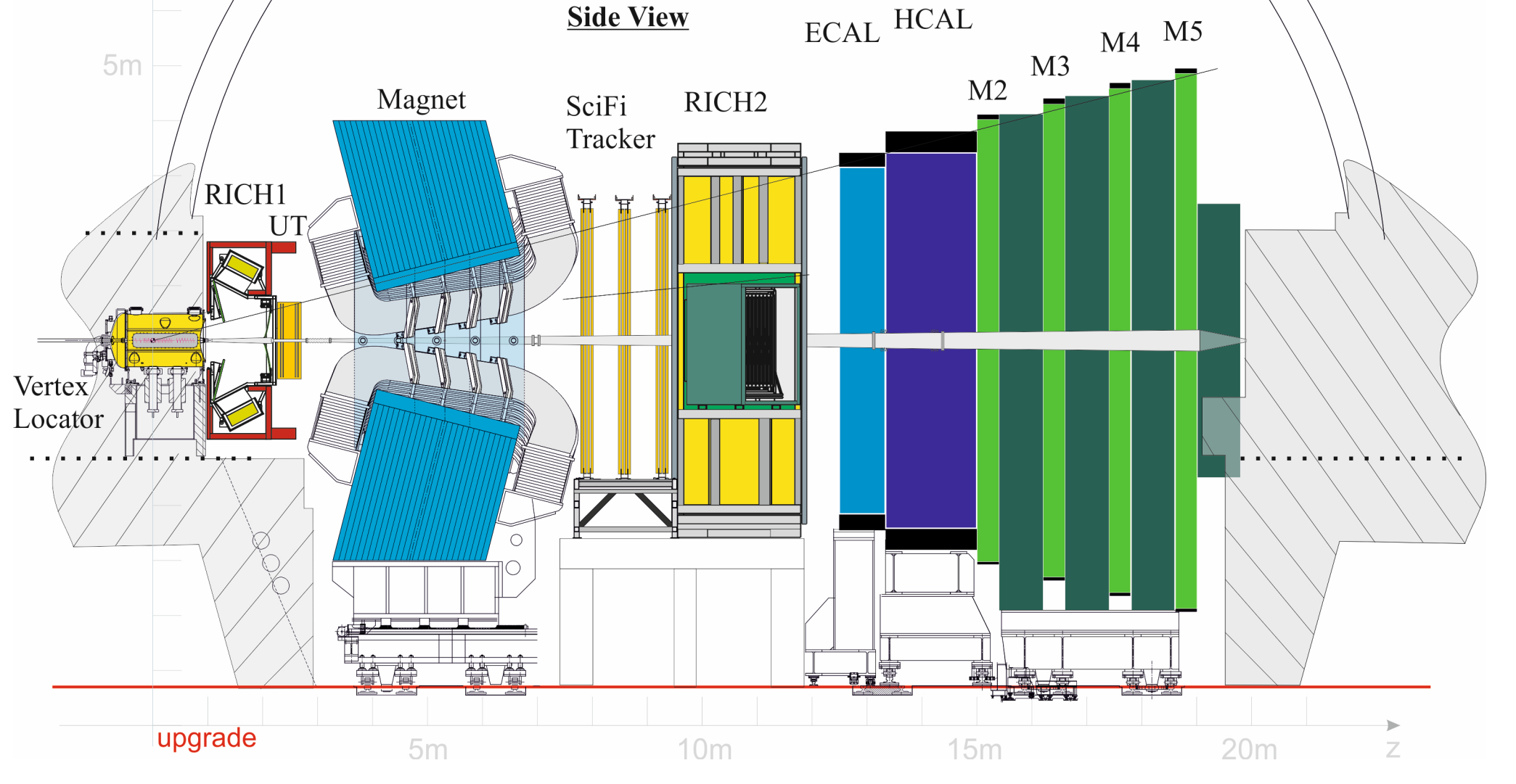}
    \caption{The new LHCb detector operating during the Run~3~\cite{LHCb:2023hlw}.}
    \label{fig:lhcb}
\end{figure}

The VErtex LOcator (VELO) is based on pixelated silicon sensors and is critical for determining the decay vertices of $b$ and $c$ flavored hadrons. The Upstream Tracker (UT) contains vertically segmented silicon strips and continues the tracking upstream of the VELO. It is also used to determine the momentum of charged particles and is useful to remove low-momentum tracks from being extrapolated downstream, thus speeding up the software trigger by about a factor of three. 
Tracking after the magnet is handled by the new scintillating fiber-based Scintillating Fiber detector (SciFi). Two Ring Imaging Cherenkov (RICH) detectors supply particle identification. RICH1 is mainly for lower momentum particles, and RICH2 is for higher momentum ones. The Electromagnetic Calorimeter (ECAL) identifies electrons and reconstructs photons and neutral pions. The Hadronic Calorimeter (HCAL) measures the energy deposits of hadrons, and four muon chambers M2-M5 are mostly used for muon identification.
The angular coverage of the LHCb detectors ranges from $ 2 < \eta < 5 $. 
Figure~\ref{fig:lhcb} shows the LHCb upgrade detector. 

\subsection{Track types at LHCb}
The tracking system of the LHCb experiment consists of three subsystems, VELO, UT, and SciFi, which are responsible for reconstructing charged particles. A magnet, with a bending power of 4~Tm, is also necessary to curve particle trajectories in order to measure their momentum, $p$. Its polarity can be inverted, and it is used to control systematic effects coming from detector inefficiencies. 

\begin{figure}[h!]
    \centering
    \includegraphics[width=0.4\textwidth]{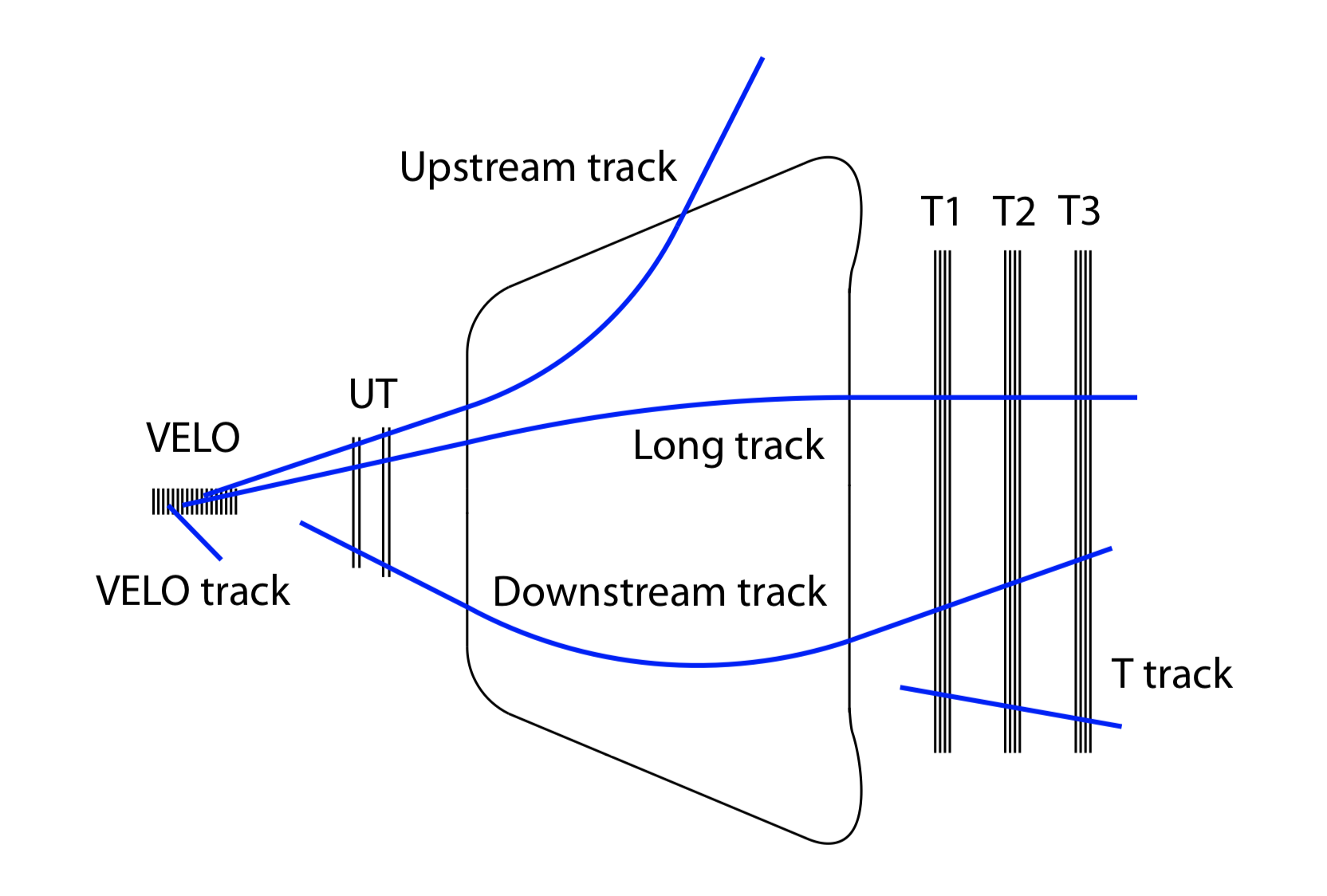}
    \caption{Definition of the particle track types in the LHCb experiment, according to which detectors are hit. The different tracker layers and the magnet in the center are sketched.}
    \label{fig:tracktypes}
\end{figure}

Several track types are defined depending on the subdetectors involved in the reconstruction, as shown in Fig.~\ref{fig:tracktypes}.

The main track types considered for physics analyses are
\begin{itemize}
    \item[ ] \underline{\textit{Long} tracks}: they have information from at least the VELO and the SciFi, and possibly the UT. These are the main tracks used in physics analyses and at all stages of the trigger;\\
    \item[ ] \underline{\textit{Downstream} tracks}: they have information from the UT and the SciFi, but not VELO. They typically correspond to decay products of \KS and \Lz hadron decays;\\
    \item[ ] \underline{\textit{T} tracks}: they only have hits from the SciFi. They are typically not included in physics analysis. Nevertheless, their potential for physics has been recently outlined \cite{arXiv:2211.10920}.
\end{itemize}

When simulating collision data, particle tracks meeting certain thresholds are defined to be {\it reconstructible} and have an assigned type according to the sub-detector reconstructibility. This is, in turn, based on the existence of reconstructed detector digits or clusters in the emulated detector, which are matched to simulated particles if the detector hits they originated from are properly linked~\cite{Li:2752971}. Requirements for \textit{long} tracks imply VELO and SciFi reconstructibility, \textit{downstream} tracks must satisfy the UT and SciFi reconstructibility, and \textit{T}-tracks only require the SciFi one.

\subsection{The High-Level Trigger (HLT)}
The trigger system of the LHCb detector in Run~3 and beyond is fully software-based for the first time. It is comprised of two levels: HLT1 and HLT2, described in detail in Ref.~\cite{CERN-LHCC-2014-016, CERN-LHCC-2020-006}.
Most notably, the HLT1 level has to be executed at a 30\,MHz rate and, as such, suffers from heavy constraints on timing for event reconstruction.

The first HLT1 trigger performs partial event reconstruction in order to reduce the data rate. Tracking algorithms play a key role in fast event decisions, and the fact that they are inherently parallelisable processes suggests a way to increase trigger performance. Thus, the HLT1 has been implemented on a number of GPUs using the \texttt{Allen} software project~\cite{Aaij:2019zbu}, which allows to manage 4\,TB/s and reduces the data rate by a factor of 30. After this initial selection, data is passed to a buffer system, which allows nearly real-time calibration and alignment of the detector. This is used for the full and improved event reconstruction carried out by HLT2.  

Due to timing constraints, the LHCb implementation in the HLT1 stage has been based on partial reconstruction and focuses solely on {\it long} tracks, i.e., tracks that have hits in the VELO. This trigger thus significantly affects the identification of particles with long lifetimes, particularly for LLP searches in LHCb, where some of the final-state particles are created further than roughly a metre away from the IP and thus outside of the VELO acceptance. A new algorithm~\cite{Jashal:2881886,LHCB-FIGURE-2023-028} has been developed and implemented to widen the reach of particle lifetimes of the HLT1 system. It is briefly described in the following.   

\subsection{The new \texttt{Downstream} algorithm}
\label{sec:downstream-algorithm}

A fast and performant algorithm has been developed to reconstruct tracks that do not let hits in the VELO detector~\cite{Jashal:2881886}.\footnote{In practice the algorithm is also performant and has a large efficiency for particles decaying after 30\,cm, being able to recover some of the \textit{long tracks} which have not been properly reconstructed.} It is based on the extrapolation of SciFi seeds (or \textit{tracklets}) to the UT detector, including the effect of the magnetic field in the $x$ coordinate. Search windows in the UT detector for hits that are compatible with tracks coming from the SciFi, and that are not used by other reconstruction algorithms, are considered. In addition, fake tracks originating from spurious hits in the detector are suppressed by a neural network with a unique hidden layer. The reconstruction efficiency for \textit{downstream} tracks of the algorithm is about 70$\%$, with ghost rates (random combinations of hits) below $20\%$. This has been verified for SM particles (\Lz and \KS) and for LLPs in the hidden sector, in the range 0.25\,GeV/c$^2$ - 4.7\,GeV/c$^2$, decaying into muons or two hadrons.   
The track momentum resolution at this stage is less than 6$\%$~\cite{LHCB-FIGURE-2023-028}, and the algorithm has a high throughput that fulfills the tight HLT1 time requirements.

\section{Signal characterisation}
\label{sec:signal_char}

\subsection{Benchmark LLP models}
Many models with LLPs exist. In this paper, some of the benchmark models recommended by the Physics Beyond Colliders (PBC) working group~\cite{Beacham:2019nyx} will be considered, with the names being \textit{BCX}:  
\begin{itemize}
    \item[1.] Dark photons $V$ (\textit{BC1}), which have kinetic mixing with $U_{Y}(1)$ SM hyperfield. Below the EW scale, the coupling is given by the kinetic mixing parameter $\epsilon$. The dark photon phenomenology (how it is produced in proton-proton collisions and its decay modes) is taken from the Refs.~\cite{SHiP:2020vbd,Ilten:2018crw}.
    \item[2.] Higgs-like dark scalars $S$. Below the EW scale $\Lambda_{\text{EW}}$, the couplings are parametrised by the $S$-Higgs mixing angle $\theta \ll 1$ and the coupling $\alpha$ of the $hSS$ operator. For \textit{BC4}, $\alpha = 0$, while for \textit{BC5}, it is fixed in a way such that $\text{Br}(h\to SS) = 0.01$. The scalar phenomenology is taken from~\cite{Boiarska:2019jym}. It is worth mentioning the difference between this description and the one used in sensitivity studies of many past experiments~\cite{Beacham:2019nyx,Antel:2023hkf}. The latter considered the so-called inclusive description of the production of the dark scalars from $B$ mesons, when the branching ratio is approximated by the process $b\to s + S$. It breaks down for large scalar masses $m_{S}\gtrsim 2-3\text{ GeV}$ (as QCD enters the non-perturbative regime, and also because of wrong scalar kinematics) and hence is inapplicable. Ref.~\cite{Boiarska:2019jym} considers the exclusive description, when the branching ratio is the sum of various decay channels $B\to \text{meson}+S$.
    \item[3.] Heavy Neutral Leptons $N$ coupled to the active neutrino $\nu_{\alpha}$: $\nu_{e}$ (\textit{BC6}), $\nu_{\mu}$ (\textit{BC7}), or $\nu_{\tau}$ (\textit{BC8}). Below the EW scale, the coupling of HNLs to the SM is via the mass mixing with active neutrinos parametrised by the HNL-neutrino mixing angle $U_{\alpha}$. The phenomenology description is taken from~\cite{Bondarenko:2018ptm}, with minor changes concerning the transition of the description of semileptonic decay widths of HNLs from the exclusive description (when the total width sums up from widths into particular meson states) to the inclusive approach (when the total width is approximated by decay into quarks).
    \item[4.] Axion-like particles (ALPs). If defined at some scale $\Lambda_{\text{ALP}}> \Lambda_{\text{EW}}$, ALPs may couple to various pseudoscalar SM operators, including Chern-Simons density of the gauge fields or the axial-vector currents of the matter; the RG dynamics down to the ALP mass scale also induces other operators. For \textit{BC10}, at $\Lambda_{\text{ALP}}$, ALPs universally couple to the fermion axial-vector current, while for \textit{BC11}, they couple to the gluon Chern-Simons density. The description of the production and decay modes of these ALPs is taken from~\cite{DallaValleGarcia:2023xhh}. Thus, the phenomenology for \textit{BC10} significantly differs from the previously adopted description of ALP production and decay modes~\cite{Beacham:2019nyx}, where many production channels and hadronic decay modes have not been taken into account. The description of decays for \textit{BC11} somewhat differs from the other study~\cite{Aloni:2018vki}, which results in a larger decay width (for the given ALP mass and coupling) and hence a smaller lifetime (see a discussion in Ref.~\cite{DallaValleGarcia:2023xhh}).
    \item[5.] $B-L$ mediator, which couples to the anomaly-free combination of the baryon and lepton currents. The coupling is given in terms of the structure constant $\alpha_{B}$. Its production and decay channels are the same as for dark photons up to the fact that the coupling is universal and there is no mixing with $\rho^{0}$ mesons~\cite{Ilten:2018crw}.
    \end{itemize}

Ref.~\cite{Ovchynnikov:2023cry} summarizes the main LLP's production and decay modes that are relevant for high-energy experiments. Mostly, they are produced directly in proton-proton collisions, decays of various SM particles, or via mixing with light neutral mesons. Therefore, most of them are relevant for LHCb. For convenience, the processes are listed in table~\ref{tab:models}.

\begin{table}[htb]
\begin{center}
\small
\begin{tabular}{|c|c|c|}
\hline
 Model & Production & Decay modes \\
\hline
Dark scalar [S] & \makecell{$B_{(s)}\to S X_s$ \\ $B\to SSX$ \\ $h\to SS$} &  \makecell{$ \ell^+\ell^-,\pi^+\pi^-, $ \\ $K^+K^-, c \bar c,  gg ... $} \\
\hline
Heavy lepton [N] & \makecell{$B/D \to N X$ \\ $W\to N+\ell$} &  \makecell{$ \ell q \bar q', \nu q \bar q $ \\ $\nu \ell \bar \ell', ... $}\\
\hline
\makecell{Massive photon [V]\\ $U_{B-L}$ mediator}& \makecell{$\pi/\eta/\eta' \to V X$\\ Bremsstrahlung \\ Drell-Yan} & \makecell{$\ell^+\ell^-,\pi^+\pi^-,$ \\ $\pi^+\pi^-\pi^0, K^+K^-$}\\
\hline
Axion-Like-Particle [a] & \makecell{$B_{(s)} \to a X_s $\\ $\pi^{0}/\eta/\eta'$ mixing \\ Drell-Yan}& \makecell{$\ell^+\ell^-,\eta2\pi,4\pi$ \\ $gg$}\\
\hline
\end{tabular} 
\end{center}
\caption{Summary of the production and decay modes of the LLPs considered in this paper. Here, $X$ denotes any SM state.}
\label{tab:models}
\end{table}

\subsection{Events selection}
\label{sec:llp-selection}

A potential event with LLPs is defined by the presence of the reconstructed decay vertex located between the end of VELO ($z\approx 1\text{ m}$) and the beginning of the UT tracker ($z_{\text{UT}}\approx 2.5\text{ m}$), in the pseudorapidity range $2<\eta<5$. 

The vertex is reconstructed with the help of at least two tracks from decay products passing through both the UT and SciFi trackers. For the present study, only charged particles are considered detectable. Therefore, decays into solely neutral particles such as $\pi^{0}(\to 2\gamma), \gamma, K^{0}_{L}$ are treated as invisible. 

\begin{figure}[h!]
    \centering
    \includegraphics[width=0.45\textwidth]{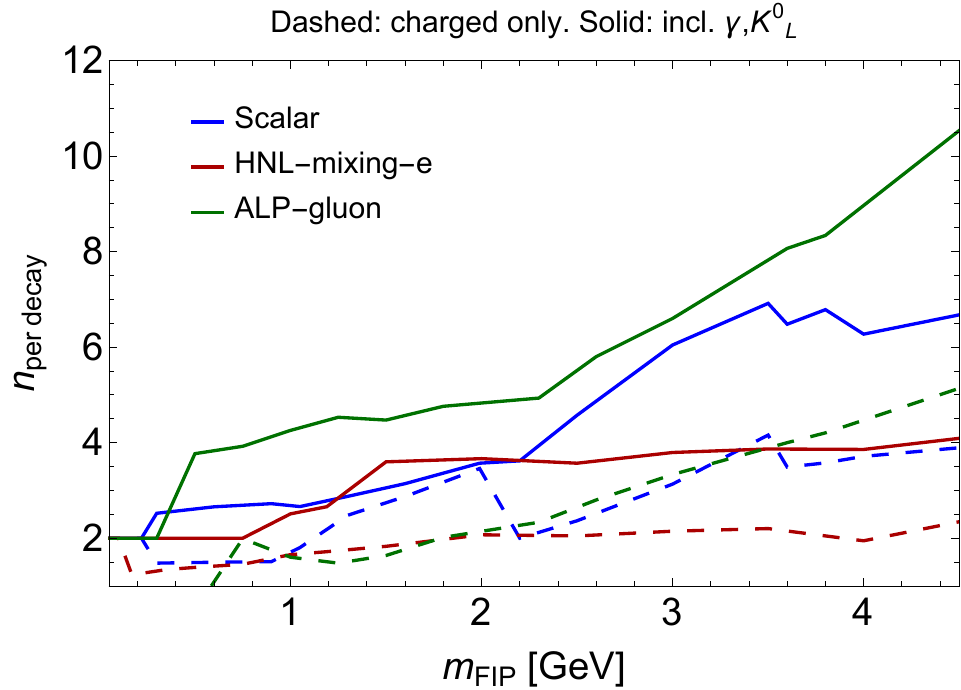}
    \caption{The average number of metastable decay products per LLP decay that may be detected -- $\pi^{\pm},K^{\pm},K^{0}_{L},\gamma,e^{\pm},\mu^{\pm}$ -- as a function of the LLP mass, for the models of HNLs coupled to the electron neutrino, Higgs-like scalars with the mixing coupling, and ALPs coupled to gluons~\cite{Ovchynnikov:2023cry}. The dashed lines assume that only charged decay products are detectable, while the solid lines also include uncharged decay products. For each case, the summation over all decay channels is performed, which may lead, in particular, to the dashed lines with $n_{\text{per decay}} < 2$ if there are modes with decays into neutral particles only. Jumps in the behavior of the lines are caused by the kinematical opening of new decay channels.}
    \label{fig:multiplicity}
\end{figure}

As indicated in Table~\ref{tab:models}, while the majority of decay modes of LLPs are exclusive two-body decays, they may often decay into three or more particles. It is especially relevant for LLPs with $m\gtrsim 1\text{ GeV}$, which decay into quarks or gluons and hence produce a cascade of hadrons resulting from showering and hadronization. Fig.~\ref{fig:multiplicity} illustrates the average multiplicity of metastable particles (those having decay lengths $c\tau p/m$ well exceeding the dimensions of LHCb) for selected models. 

This feature necessitates a consistent approach to LLP reconstruction. Reconstructing the many-particle vertex by as few tracks as possible clearly maximizes the yield of reconstructed events. Namely, each track is reconstructed with finite efficiency, which results from the non-ideal performance of the detector, which introduces a finite detection efficiency and kinematics measurement resolution. However, reconstructing more particles from the vertex and using PID criteria,\footnote{At HLT1 level, information from the electromagnetic calorimeter and muon chambers is available. Efforts are also being performed to include the information of the RICH detectors.} one may reveal the properties of the LLP and hence discern different LLP scenarios (see, e.g.,~\cite{Mikulenko:2023}).

For the present study, the main interest is in estimating the region of LLP's parameter space where the \texttt{Downstream} algorithm may see any signal. In this sense, it is enough to have two reconstructed tracks. The event reconstruction efficiency is then approximated by the squared reconstruction efficiency of the single track times the vertex reconstruction efficiency. The opportunities of the reconstruction by using many tracks will be studied in the future.

The event reconstruction performance of the \texttt{Downstream} algorithm is a subject of ongoing investigation. It includes, e.g., the momentum dependence of the track reconstruction efficiency and the two-\textit{downstream}-track vertex resolution.\footnote{The latter is expected to degrade with its mass $m$. However, the amount of background is expected to decrease in the domain of larger $m$, thereby rendering a larger mass resolution for high LLP masses less likely to impact the searches significantly.} For the reference selection in this paper, the particles will be required to have the energy $E>5\text{ GeV}$, and transverse momentum $p_T>0.5\text{ GeV}$, 
and the overall event reconstruction efficiency, $\epsilon_{\text{rec}} = 0.4$ is considered. 

Potential ways to enhance the event yield that will be studied in the future are worth mentioning. First, it may be significantly improved if extending the $z$ range with the reconstructed vertex until the beginning of the first SciFi layer, which is located at $z \approx 7.7\text{ m}$. Then, the vertices from $z >z_{\text{UT}}$ would be reconstructed with the help of the SciFi tracker only (i.e., using solely $T$ tracks). Second, a sizable fraction of decays of LLPs may be into neutral particles such as $\gamma$ and $K^{0}_{L}$. Some particles, such as light ALPs coupled to gluons and ALPs coupled to photons, decay solely into photons. Therefore, adding the option of reconstructing events using calorimeters would be essential for these LLPs.

\subsection{Case study: dark scalars}

\begin{figure}
    \includegraphics[width=0.45\textwidth]{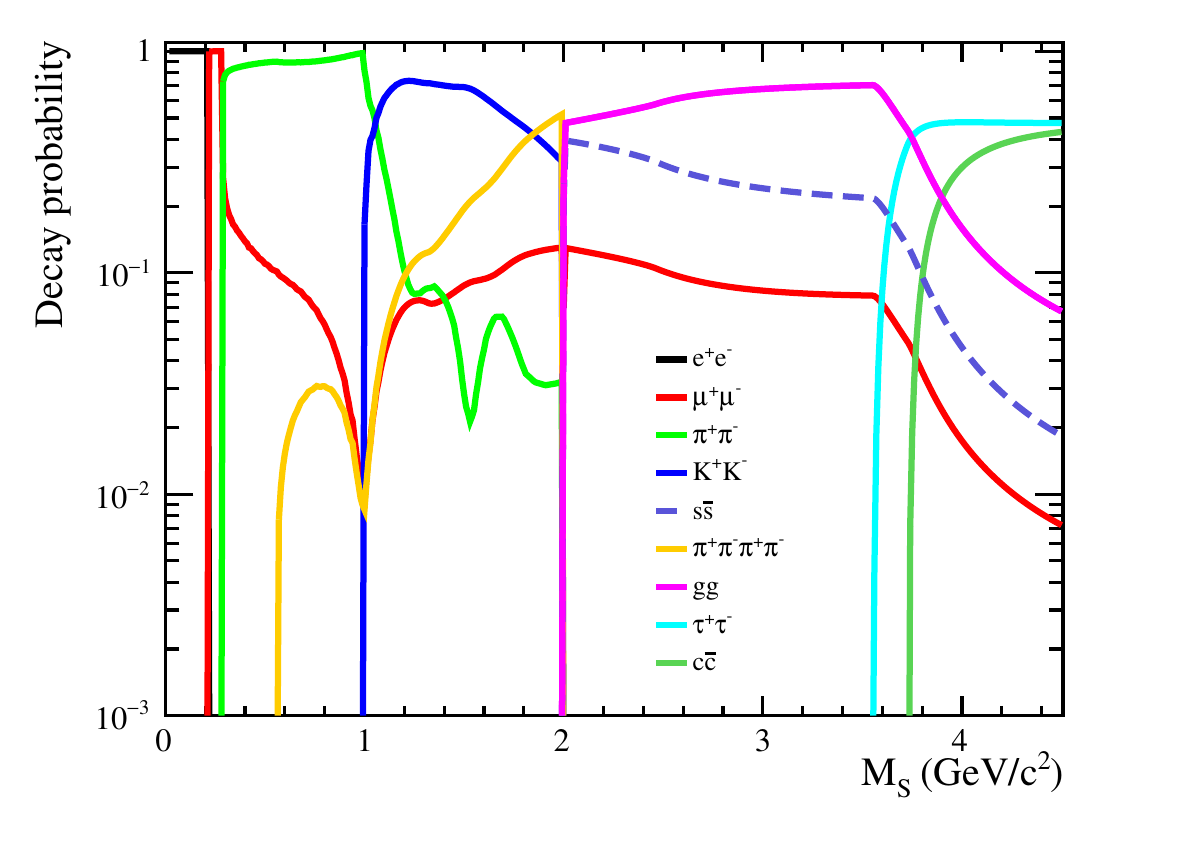}
    \caption{Decay probabilities of a dark scalar into different channels as a function of its mass and normalised to unity~\cite{Boiarska:2019jym}.}
\label{fig:decays}
\end{figure}

Of particular interest is the dark scalar model denoted as \textit{BC4}. These scalars can be generated through processes such as $B\to S+X_{s/d}$, where $X_{q}$ denotes a hadronic state containing the quark $q$. For $m_{S}\ll m_{B}$ and in the limit $\theta^{2} \ll 1$, the collective branching ratio for these processes is of the order of $3.3\theta^{2}$, the production threshold is, approximately, $m_B - m_\pi\approx 5.13$ GeV$/c^2$~\cite{Boiarska:2019jym}. Fig.~\ref{fig:decays} illustrates the scalar's decay probabilities as a function of its mass, normalized to unity.

Decays involving two muons and electrons are particularly pertinent for particles with masses below 1\,GeV$/c^2$, while the $\pi\pi/KK$ channels dominate within the 0.270-2\,GeV$/c^2$ mass range. From a mass threshold of 2\,GeV$/c^2$ onward, there is a proliferation in track multiplicity, coinciding with the opening of various channels such as gluon-gluon ($gg$), $s\bar s$, $c\bar c$, and $\tau^+\tau^-$. These channels assume particular importance due to the expectation of three or more ``downstream'' tracks originating from a common vertex. In the case of the $c\bar c$ decay channel, two $D$ mesons and many pions will be produced as a result of showering and hadronization. The $D$s will decay afterward, and formally, the event would be a bunch of soft hadrons from the LLP's decay vertex and two displaced hadron showers from $D$s decays. However, the magnitude of the displacement, proportional to the decay length of $D$ mesons, is well below the vertex resolution, so all the tracks should converge to the same origin.

\section{Background sources}
\label{sec:backgrounds}

Background events that could mimic the BSM signal at LHCb are expected to arise from different sources~\cite{LHCb:2016awg}. They are listed below. Some of these background events can be studied with simulations \cite{Mazurek:2022tlu}, and other sources will be studied when Run~3 data is available.  
The main contributions are considered to come from: 

\begin{itemize} 
\item[$\bullet$] Hadronic resonances: decays of light and heavy $q\bar{q}$ resonances into a pair of hadrons ($h^+h^-$) or leptons ($\ell^+\ell^-$) are highly suppressed since they decay promptly and from simulation studies no tracks are expected to be reconstructible as \textit{downstream} tracks, neither if they come from the interaction point nor from decays of $b$ and $c$ hadron decays.
Light resonances can be produced by particle interaction with the beam pipe or detector material, decaying into muons or pions. This background can be suppressed by using control samples from data and vetoing specific regions of the detector. 

\item[$\bullet$] Strange candidates: SM particles with long lifetimes (notably \KS and \Lz) can also be mistaken as signal events. This could happen when the LLP is reconstructed in hadronic $h^+h^-$ modes or for leptonic modes if the hadrons from the \KS, or the proton and pion from the \Lz, are misidentified as muons~\footnote{Decays of \KS to two leptons are highly suppressed in the SM, with branching fractions of order $10^{-12}$.}. This type of background can be rejected by imposing tighter particle identification (PID) criteria and by vetoing pairs of particles that, after being assigned the proton or pion mass hypothesis, lie in the invariant mass region of \KS and \Lz candidates.

\item[$\bullet$] Combinatorial background: random pairs of hadrons or leptons, associated or not with other particles from $B$-meson decays, could be wrongly attributed to LLP candidates. MC simulations show that the amount of combinatorial background drastically decreases with the mass of the LLP particle, being negligible for masses larger than 2 \,GeV. This is expected since high momentum tracks come from decays of $b$ and charm hadrons. Information on the two-tracks and $B$-meson candidates can be used in a multivariate analysis, in particular, making use of a boosted decision tree (BDT) or neural network (NN), which are very suitable to reduce this source of background. The vertex quality, impact parameter, transverse momenta, or track isolation criteria are examples of variables that are expected to be very discriminant in this type of analysis.
\end{itemize} 

A NN classifier can be used to suppress the background events, with a threshold that can be varied according to the desired performance. Using simulated events~\cite{Mazurek:2022tlu}, a background rejection rate larger than 99$\%$ and a signal efficiency of 87$\%$ can be obtained, assuming two-body decays for the latter. In this test the NN is trained using dedicated signal samples with BSM candidates, in particular using dark scalars with masses ranging between 400~MeV and 4500~MeV. Background events are obtained from minimum bias simulations\footnote{Collisions that occur without any specific selection criteria applied.}. Input variables are track properties of the reconstructed pairs (impact parameter, momentum and transverse momentum), vertex quality and position, and impact parameter, quality, and momentum of the reconstructed parent particle.

This background reduction is expected since, at large lifetimes, most of the background is coming from material interaction, which has a very different topology and kinematics than the signal. The rejection rate could be even higher if the LLP decays into multiple particles.

Secondary interactions of hadrons produced in beam-gas collisions can be used to map the location of material as it is done in Ref.~\cite{Alexander:2018png}. With this procedure, the background can be reduced to a negligible level.   

\section{LLP events yield and qualitative comparison with other proposals}
\subsection{Signal yield}
\label{sec:signal}

The LLP exploration power of the \texttt{Downstream} algorithm is estimated in the following.  

To calculate the number of events with LLPs, it is necessary to know their production channels, the fraction of LLP flying in the direction of the detector, the decay probability, and the fraction of the decay events that may be reconstructed. The semi-analytic approach described in~\cite{Ovchynnikov:2023cry,Bondarenko:2019yob} is used, which may be as accurate as pure Monte-Carlo evaluation, combining this with transparency and speed of calculations. The number of events is calculated as
\begin{multline}
    N_{\text{ev}} = \mathcal{L}\sum_{i}\sigma^{(i)}_{pp\to \text{LLP}} \, \int d\theta dE dz \, f^{(i)}(\theta,E)\cdot\epsilon_{\text{az}}(\theta,z) \times \\ \frac{dP_{\text{dec}}}{dz}\cdot\epsilon_{\text{det}}(m,\theta,E,z)\cdot \epsilon_{\text{rec}}\cdot \epsilon_{\text{S/B}}
    \label{eq:Nevents}
\end{multline}
The quantities entering Eq.~\eqref{eq:Nevents} are the following:
\begin{itemize}
\item[--] $\mathcal{L}$ is the total integrated luminosity corresponding to the operating time of the experiment.
\item[--] $\sigma_{pp\to \text{LLP}}^{(i)}$ is the LLP cross section in proton-proton collisions, accounting for the probability that a specific process $i$ takes place, e.g., decays of mesons, direct production by proton-target collisions, etc.
\item[--] $z$, $\theta$, and $E$ are, respectively, the position along the beam axis, the polar angle, and the energy of the LLP.
\item[--] $f^{(i)}(\theta,E)$ is the differential distribution of LLPs produced in the process $i$ in polar angle and energy.
\item[--]$\epsilon_{\text{az}}(\theta,z)$ is the azimuthal acceptance: 
    \begin{equation}
    \epsilon_{\text{az}} = \frac{\Delta \phi_{\text{decay volume}}(\theta,z)}{2\pi}
    \end{equation}
where $\Delta\phi$ is the fraction of azimuthal coverage for which LLPs decaying at $(z,\theta)$ are inside the decay volume. For the specified setup, $\epsilon_{\text{az}} = h(2<\eta(\theta)<5)$, where $h$ is the step function. 
\item[--] $\frac{dP_{\text{dec}}}{dz}$ is the differential decay probability: 
\begin{equation}
\frac{dP_{\text{dec}}}{dz} = \frac{\exp[-r(z,\theta)/l_{\text{dec}}]}{l_{\text{dec}}} \frac{dr(z,\theta)}{dz},
\label{eq:decay-probability}
\end{equation}
with $r = z/\cos(\theta)$ being the modulus of the displacement of the LLP decay position from its production point, and $l_{\text{dec}}=c\tau \sqrt{\gamma^2-1}$ is the LLP decay length in the lab frame (with $\tau$ being the lifetime in terms of the LLP mass and the coupling to the SM particles $g$).
\item[--] $\epsilon_{\text{det}}(m,\theta,E,z)$ is the decay products acceptance, i.e., among those LLPs that are within the azimuthal acceptance, the fraction of LLPs that have at least two decay products that point to the detector and that may be reconstructed. Schematically,
\begin{equation}
    \epsilon_{\text{det}} = \sum_{j}\text{Br}_{\text{vis}}^{(j)}(m)\cdot \epsilon_{\text{det}}^{\text{(geom,j)}}\cdot \epsilon_{\text{det}}^{\text{(other cuts,j)}},
    \label{eq:decay-acceptance-schematic}
\end{equation}
where $j$ counts over the LLP decays into final states (with the branching ratio denoted as $\text{Br}_{\text{vis}}$) that are detectable. Depending on the presence of a calorimeter (EM and/or hadronic), they may encompass only those states featuring at least two charged particles, or also (if the calorimeters are present) the states with at least two neutral particles. For the \texttt{Downstream} algorithm, only charged decay products are considered as visible; this way, the acceptance estimates are conservative. Generically, the reconstructed decay may also include some neutral states such as photons and $K^{0}_{L}$. $\epsilon_{\text{det}}^{\text{(geom)}}$ denotes the fraction of visible decay products that point to the end of the detector (which is SciFi in the case of the \texttt{Downstream} setup), and $\epsilon_{\text{det}}^{\text{(other cuts)}}$ is the fraction of these decay products that additionally satisfy the remaining selection criteria (e.g., minimum energy requirement, etc.).
\item[--]$\epsilon_{\text{rec}} \approx 0.4$ is the reconstruction efficiency, i.e., the fraction of the events that pass the azimuthal and decay acceptances criteria that the detector can successfully reconstruct (remind Sec.~\ref{sec:llp-selection}). 
\item[--] Finally, $\epsilon_{\text{S/B}}$ is the signal-preserving efficiency for the events that have been reconstructed, resulting from the background rejection. This efficiency is assumed to be $87\%$ on average.
\end{itemize}

The number of events is calculated using Eq.~\eqref{eq:Nevents}, the \texttt{Downstream} setup in \texttt{SensCalc} code~\cite{Ovchynnikov:2023cry} is incorporated. 
A detailed discussion on the implementation and its validation by the comparison with the LHCb simulation framework can be found in Appendix~\ref{app:senscalc-validation}.  In Table~\ref{tab:comparison-parameters}, the parameters of the setup used for the implementation are described.

Here and below, it is assumed that the search will be performed in the regime when the background is negligible, resulting from a high performance of the signal selection criteria using neural network techniques.

\subsection{Comparison with LHC-based experiments}
\label{sec:comparison-lhc-based}
To understand the LLP exploration abilities of the new \texttt{Downstream} algorithm, it is necessary to compare the LLP event yields at LHCb with LHC-based experiments. For the reference cases of the latter, the FASER and FASER2 experiments~\cite{FASER:2018bac,FASER:2018ceo} are considered. FASER, a Forward Search Experiment at the LHC designed to study neutrinos and search for weakly interacting, light new particles, is a currently running experiment located 480 m downward the ATLAS interaction point, in the far-forward direction. FASER2 is a possible upgrade of FASER with increased geometric size. It may either be located at the same placement as FASER, or at the Forward Physics Facility~\cite{Feng:2022inv}; the first setup is considered here. Apart from the fact that FASER is already running, this choice is motivated by the fact that FASER and FASER2 have the same capabilities in reconstructing the LLP kinematics (such as measuring the invariant mass and identifying the decay products) as LHCb. The operating time of FASER is LHC Run 3, while for FASER2, it is HL-LHC.  

\begin{table*}[t!]
    \centering
    \begin{tabular}{|c|c|c|c|c|c|c|c|c|}
     \hline Experiment & $\mathcal{L} (\text{ fb}^{-1}$) & $(z_{\text{min}},z_{\text{max}}) (\text{ m})$ & $(\theta_{\text{min}},\theta_{\text{max}}) (\text{ mrad})$  & Selection \\ \hline
      LHCb with \texttt{Downstream} & \makecell{25 (w FASER)\\ 300 (w FASER2)} & (1,1.5) & $\approx(1.3,260)$ & \makecell{Two oppositely charged particles
       \\ enter SciFi, $p > 5\text{ GeV}/c$ \\ $\epsilon_{\text{rec}}\approx 0.4$}  \\ \hline
      FASER & 150 & $(480,481.5)$ & $\approx (0,0.21)$ & \makecell{Two particles with zero total charge \\ intersect the detector}  \\ \hline
       FASER2  & 3000 & (480,485) & $\approx (0,2.1)$& \makecell{Two particles with zero total charge \\ intersect the detector} \\ \hline
    \end{tabular}
    \caption{Setups of the LHCb with the \texttt{Downstream} algorithm and the FASER and FASER2 experiments used for the comparison of the signal rates. The columns are: the name of the experiment, the integrated luminosity, the minimal and maximal longitudinal displacement covered by the decay volume, the minimal and maximal angles covered by the decay volume, and the selection criteria imposed on the LLP decay. Two different luminosities are considered for the \texttt{Downstream} algorithm in order to make a proper comparison with FASER/FASER2 (see text for details).}
    \label{tab:comparison-parameters}
\end{table*}

The list of the relevant parameters of the considered experiments is given in Table~\ref{tab:comparison-parameters}. For the LHCb experiment with the new \texttt{Downstream} algorithm a partial statistics of Run 3, $\mathcal{L} = 25\text{ fb}^{-1}$, is considered when comparing with FASER\footnote{For the LHCb experiment an integrated luminosity $\mathcal{L} = 15\text{ fb}^{-1}$ is expected for 2024, and a minimum of $\mathcal{L} = 30\text{ fb}^{-1}$ for the full Run3.} and the full statistics until Run 6 $\mathcal{L} = 300\text{ fb}^{-1}$ are assumed for LHCb when comparing with FASER2. A conservative configuration of the LHCb setup is considered with the effective decay volume from $z = 1\text{ m}$ (the end of VELO) and until the UT layers. For the \texttt{Downstream} algorithm, it is required that the charged decay products have $E>5\text{ GeV}$. For FASER and FASER2, the setups implemented in \texttt{SensCalc} are used without the requirement of any other selection criteria than the requisite for the decay products to pass through the detector.

Considering the limit when $c\tau \langle \gamma\rangle \gg \Delta X_{\text{exp}}$, where $\Delta X_{\text{exp}}$ is the geometric size of the whole experiment, from the production point and until the end of the detector, the differential decay probability~\eqref{eq:decay-probability} reduces to $dP_{\text{dec}}/dz \approx 1/(l_{\text{dec}}\cos(\theta))$. The expression~\eqref{eq:Nevents} becomes
\begin{equation}
    N_{\text{ev,lower}} \approx 
    \mathcal{L}\sum_{i}\sigma^{(i)}_{pp\to\text{LLP}}\cdot \epsilon^{(i)},
    \label{eq:number-of-events-large-lifetimes}
\end{equation}
where $\epsilon^{(i)}$ is the total acceptance for the given production channel:
\begin{equation}
    \epsilon^{(i)} = \int d\theta dE dz\ f^{(i)}\cdot \epsilon_{\text{az}}\cdot \frac{\epsilon_{\text{det}}}{\cos(\theta)c\tau\sqrt{\gamma^{2}-1}}
    \label{eq:total-acceptance}
\end{equation}
This quantity may be decomposed as
\begin{equation}
    \epsilon = \langle \epsilon_{\text{LLP}}\rangle\cdot \frac{\Delta z}{c\tau} \langle (\gamma^{2}-1)^{-1/2}\rangle\cdot \langle \epsilon_{\text{det}}\rangle,
\end{equation}
where $\langle \epsilon_{\text{LLP}}\rangle$ is a fraction of LLPs that intersect the decay volume, $\Delta z = 1.5\text{ m}$ is the longitudinal length of the effective decay volume, $\frac{\Delta z}{c\tau}\langle (\gamma^{2}-1)^{-1/2}\rangle$ is the mean decay probability for the LLPs intersecting the decay volume, and $\langle \epsilon_{\text{det}}\rangle$ is the mean decay products acceptance for the LLPs decayed inside. The Eq.~\eqref{eq:number-of-events-large-lifetimes} is very convenient for the comparison since the dependence on the LLP lifetime factorizes out. In particular, given the coupling $g$ of the LLP to the SM, the minimal possible value of $g$ that may be probed is given by
\begin{equation}
g^{2}_{\text{lower}}(m) \approx (N_{\text{ev,lower}}|_{g= 1})^{-1/2},
\end{equation}
which follows from the scaling $N^{(i)}_{\text{prod}},\tau^{-1}\propto g^{2}$ in Eq.~\eqref{eq:number-of-events-large-lifetimes}.

\begin{figure*}[t!]
    \centering
    \includegraphics[width=0.45\textwidth]{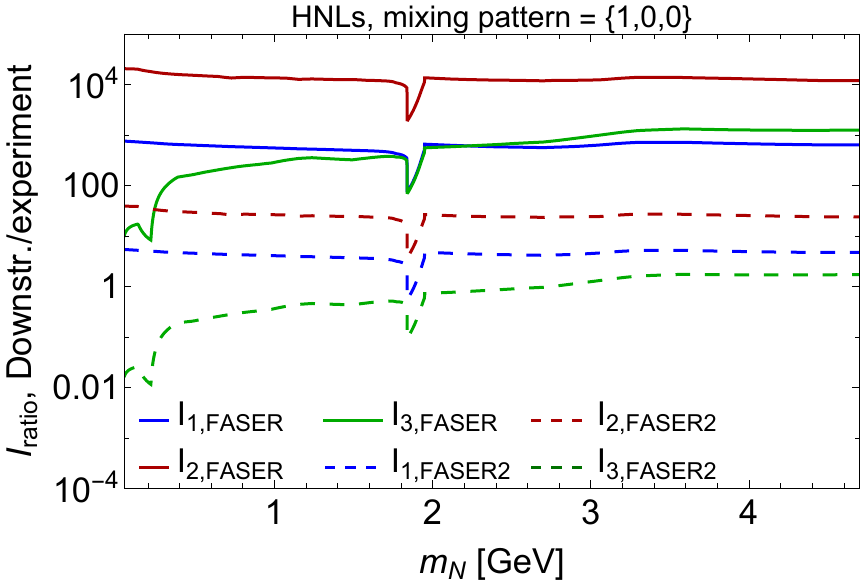}~\includegraphics[width=0.45\textwidth]{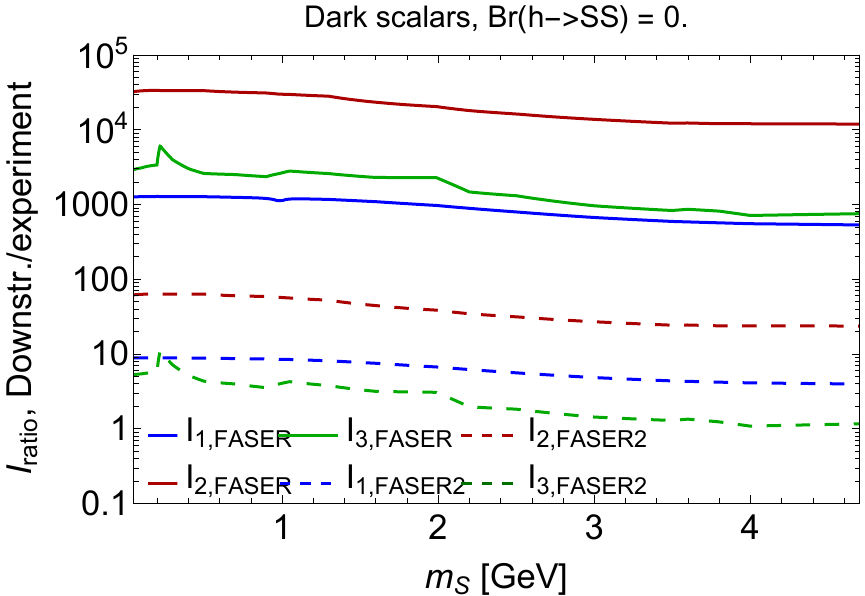}
    \caption{The ratio of the quantities $\mathcal{I}_{i}$ (see the text for definition) for the events at \texttt{LHCb-Downstream} and FASER (solid lines) or FASER2 (dashed lines) for the models of heavy neutral leptons mixing with $\nu_{e}$ (the left panel) and dark scalar mixing with the Higgs boson (the right panel). The ratios have been computed using \texttt{SensCalc}~\cite{Ovchynnikov:2023cry}.}
    \label{fig:acceptances-comparison}
\end{figure*}

To understand the impact of different luminosities, angular coverage, and decay volume length, in Eq.~\eqref{eq:total-acceptance}, the setting $\epsilon = 1$ is first applied, and then the various factors entering in the integrand sequentially included. The quantities that are compared are: $\mathcal{I}_{0}$ -- the total number of LLPs produced during the runtime of the experiment ($\epsilon = 1$); $\mathcal{I}_{1}$ -- the fraction of LLPs pointing to the decay volume (only $f^{(i)}\epsilon_{\text{az}}/\Delta z$ is included in Eq.~\eqref{eq:total-acceptance}); $\mathcal{I}_{2}$ -- the fraction of the LLPs decaying inside (all the factors except for $\epsilon_{\text{det}}\cdot \epsilon_{\text{rec}}\cdot \epsilon_{S/B}$ are included); $\mathcal{I}_{3}$ -- the fraction of the decay events which pass the reconstruction (all the factors are included). 

In Fig.~\ref{fig:acceptances-comparison}, the expression to obtain the number of events~\eqref{eq:number-of-events-large-lifetimes} for the model of dark scalars and heavy neutral leptons coupled to the electron flavor are compared. Decays of $B$ and $D$ (for HNLs) mesons produce these particles, while their visible decays are leptonic, hadronic (for scalars), or semileptonic 
(for HNLs)~\cite{Boiarska:2019jym,Bondarenko:2018ptm}. Because of the similar proton collision energy, the only difference in $\mathcal{I}_{0}$ comes from different integrated luminosities accumulated during the runtime of the experiments; the ratio is constant and equal to $\mathcal{I}_{0,\text{Downstr}}/\mathcal{I}_{0,\text{FASER/FASER2}} \approx 0.08/0.1$ for the two luminosity values that are considered (and we do now show it in the plot). These are larger for FASER and especially for FASER2. 
However, a smaller angular coverage of the latter experiments means that a much smaller fraction of the produced particles would fly to the decay volume ($\mathcal{I}_{1}$). The decay probability approximately scales as $\Delta z\cdot \langle p^{-1}\rangle$. Overall, this ratio is much smaller at FASER and FASER2 experiments: the LLPs flying in the far-forward direction have mean momenta $\mathcal{O}(1 \text{ TeV}/c)$, while LLPs within the angular coverage of LHCb typically have $p\sim 50-100\text{ GeV}/c$. Including the decay products acceptance $\epsilon$ does not lead to a qualitative change in the ratio of the number of events, especially if most of the decay modes contain at least two charged particles. In the case when there are only uncharged particles, it is conservatively assumed that it is not possible to reconstruct them with the \texttt{Downstream} algorithm, while FASER/FASER2 are equipped with the calorimeter and hence may reconstruct such modes.

Moreover, allowing the LLPs to decay between the UT and the SciFi layers, with the reconstruction of \textit{faraway} tracks ($T$-tracks), will increase the decay probability even further.

It is also useful to compare the sensitivity to ``short-lived'' LPPs, i.e., to those for which the typical decay length is similar to the distance to the decay volume $z_{\text{min}}$, $c\tau\langle p\rangle/m \lesssim z_{\text{min}}$. In this case, the scaling of the number of events with $g$ is mainly due to the exponentially suppressed decay probability $P_{\text{dec}}\approx \exp[-z_{\text{min}} m/c\tau p]$. The scaling of the maximal value of the probed $g$ may be roughly estimated as $g_{\text{upper}} \propto \langle p\rangle/z_{\text{min}}$~\cite{Bondarenko:2019yob}. Taking into account that the LLPs at FASER/FASER2 and LHCb have the momenta of the order of $1\text{ TeV}/c$ and $100\text{ GeV}/c$ correspondingly, and using $z_{\text{min}}$ from Table~\ref{tab:comparison-parameters}   
\begin{equation}\frac{g_{\text{upper}}^{\text{Downstream}}}{g_{\text{upper}}^{\text{FASER/FASER2}}} \sim 50
\label{eq:upper-bound-faser}
\end{equation}
is obtained. 

To summarize, for the exploration power of extremely long-lived particles, the LHCb experiment with the inclusion of the new \texttt{Downstream} algorithm would perform much better than FASER and comparable to FASER2. In the parameter space where LLPs are short-lived, such that they decay before reaching the decay volume, the algorithm would deliver a better sensitivity because of a much smaller distance to the decay volume.

\section{Sensitivity to LLPs}
\label{sec:sensitivity}

To estimate the sensitivity, it is required $N_{\text{events}}>2.3$, which corresponds to the 90\% CL limit if assuming that the background is negligible~\cite{Feldman:1997qc,Bityukov:2000ms} (remind Sec.~\ref{sec:backgrounds}). Two values of the integrated luminosities will be considered: $\mathcal{L} = 25\text{ fb}^{-1}$, corresponding to partial statistics accumulated during Run3 with the \texttt{Downstream} algorithm available, and $\mathcal{L} = 300\text{ fb}^{-1}$, corresponding to the full HL-LHC phase.

\begin{figure*}[t!]
    \centering
    \includegraphics[width=0.45\textwidth]{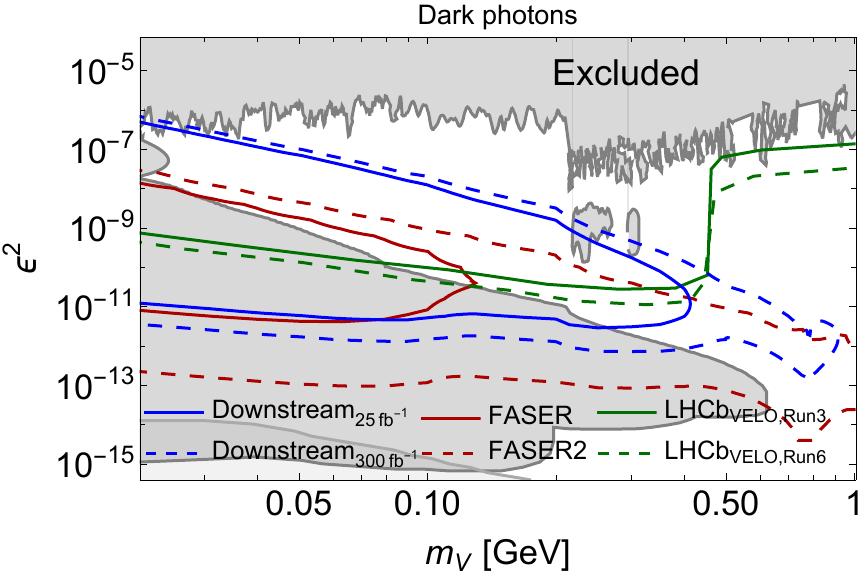}~\includegraphics[width=0.45\textwidth]{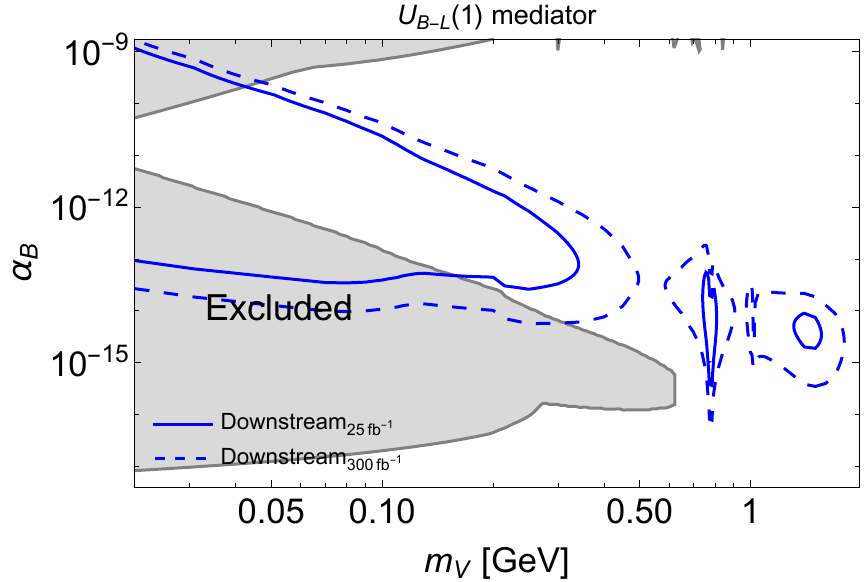}
    \caption{Sensitivity to dark photons (\textit{BC1}, the left panel) and $B-L$ mediators (the right panel) in the plane LLP mass-LLP coupling. The sensitivity of future LHCb searches restricted by VELO is taken from~\cite{LHCbCollaboration:2806113}, while the excluded parameter space and the sensitivity of FASER and FASER2 experiments is taken from~\cite{Antel:2023hkf}. For the \texttt{Downstream algorithm}, in this and subsequent figures, two values of the integrated luminosity are assumed: $25\text{ fb}^{-1}$, corresponding to the partial statistics of Run 3, and $300\text{ fb}^{-1}$, which is the full statistics of Run 6. For the description of the models, see Sec.~\ref{sec:signal_char} and Ref.~\cite{Ovchynnikov:2023cry}. See the text for the discussion on the sensitivity.}
    \label{fig:sensitivities-DP}
\end{figure*}

\begin{figure*}[t!]
    \centering
    \includegraphics[width=0.45\textwidth]{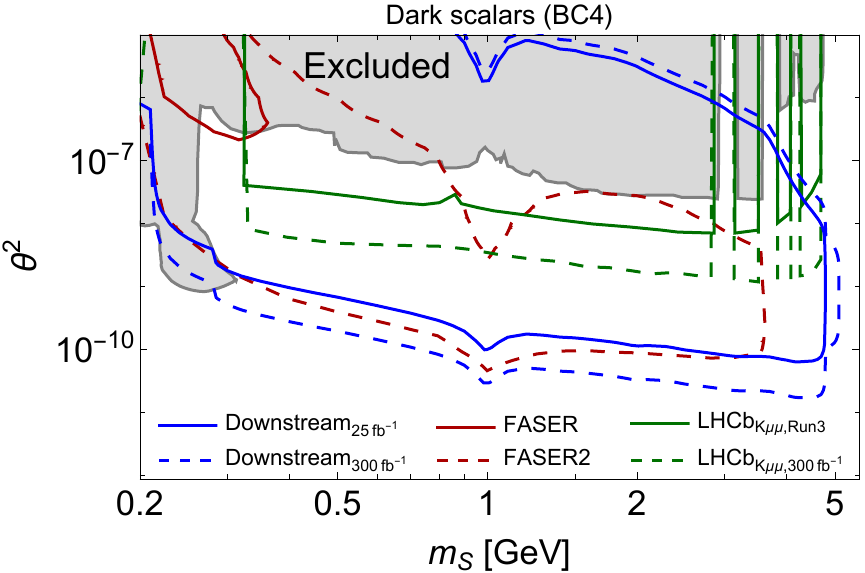}~\includegraphics[width=0.45\textwidth]{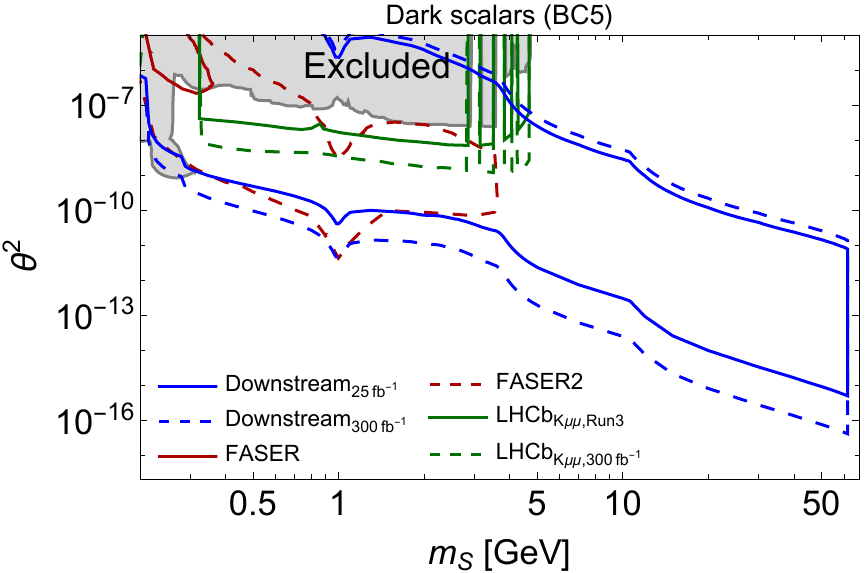}
    \caption{Sensitivity to Higgs-like scalars, models \textit{BC4} (the left panel) and \textit{BC5} (the right panel). The excluded domain, as well as sensitivities of FASER, FASER2, and the search of $B\to KS(\to \mu\mu)$ are taken from~\cite{Antel:2023hkf}.}
    \label{fig:sensitivities-Scalar}
\end{figure*}

\begin{figure*}[t!]
    \centering
    \includegraphics[width=0.45\textwidth]{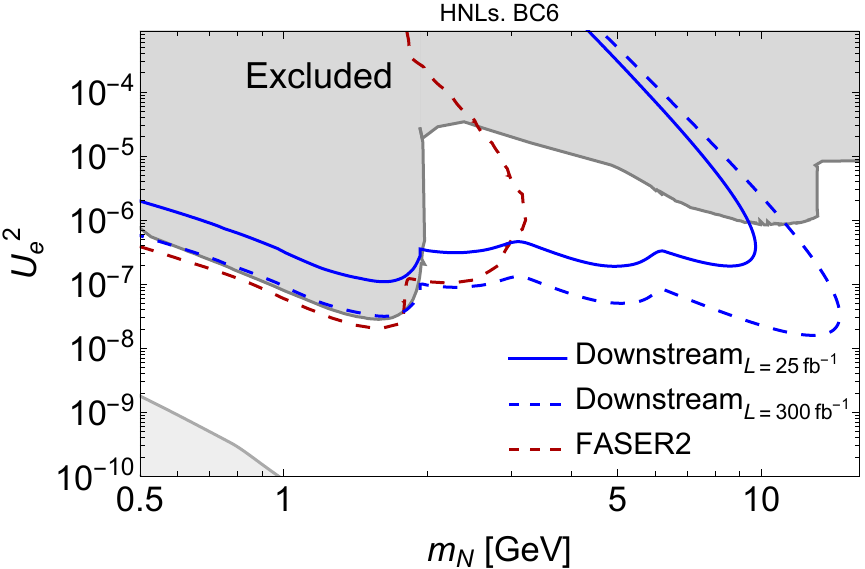}~\includegraphics[width=0.45\textwidth]{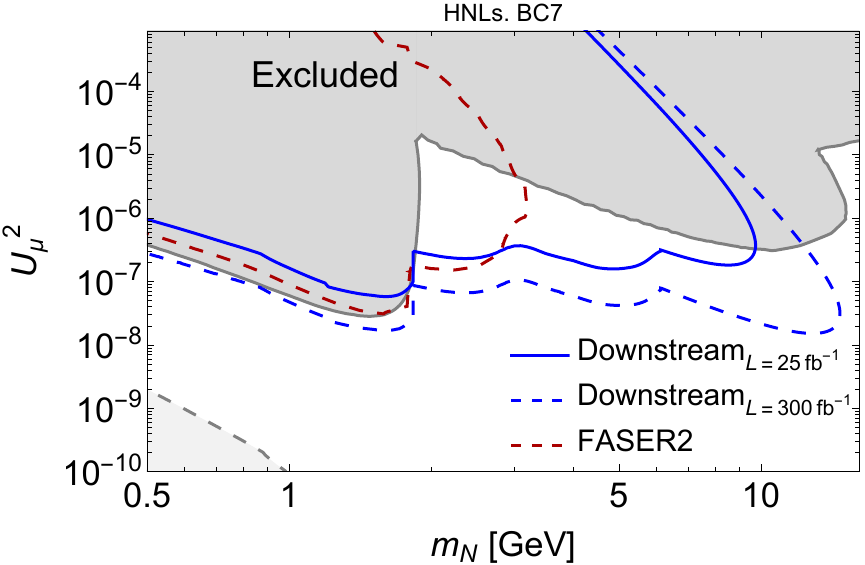}
    \\ \includegraphics[width=0.45\textwidth]{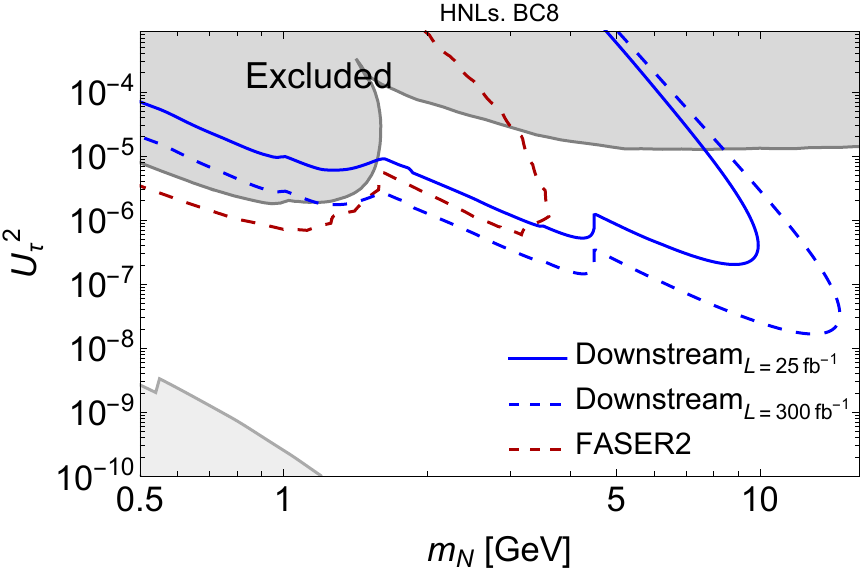}
    \caption{Sensitivity to HNLs coupled solely to $\nu_{e}$ (the top left panel), $\nu_{\mu}$ (the top right panel), and $\nu_{\tau}$ (the bottom panel). The parameter space excluded by past experiments, as well as the sensitivity of FASER2, are taken from~\cite{Antel:2023hkf}. The bottom gray domain below the short-dashed line corresponds to the parameter space excluded by BBN~\cite{Boyarsky:2020dzc,Sabti:2020yrt}.}
    \label{fig:sensitivities-HNL}
\end{figure*}

\begin{figure*}[t!]
    \centering
    \includegraphics[width=0.45\textwidth]{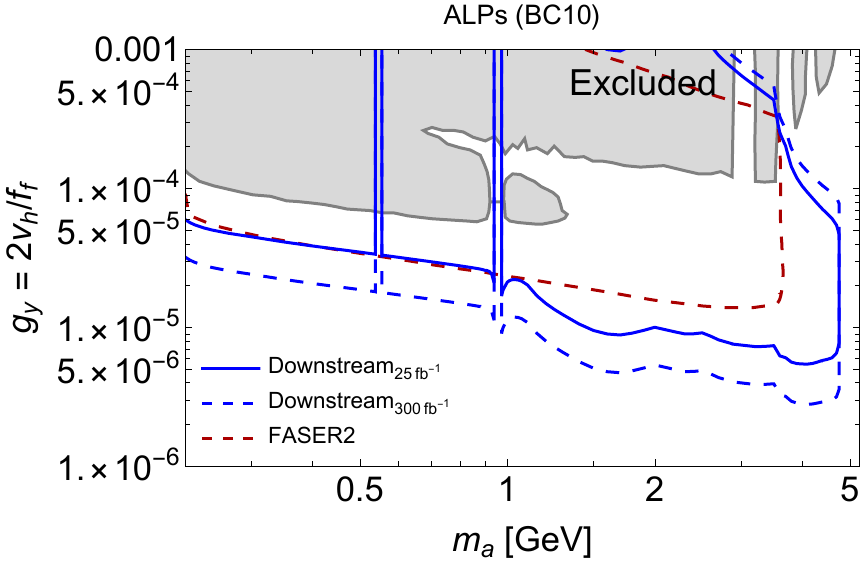}~\includegraphics[width=0.45\textwidth]{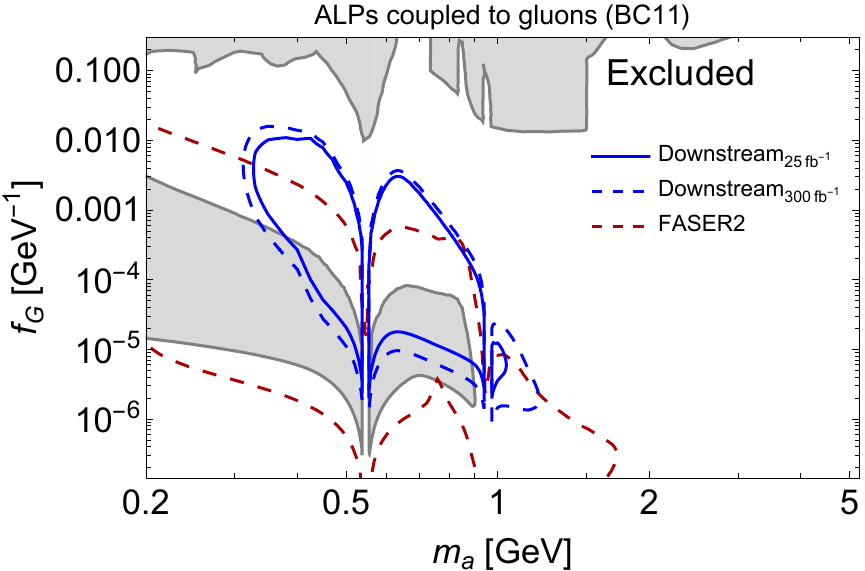}
    \caption{The sensitivity to the ALPs universally coupled to fermions (\textit{BC10}, the left panel) and to gluons (\textit{BC11}, the right panel). The sensitivity of FASER2 and the excluded parameter space are taken from~\cite{Antel:2023hkf}. For the discussion of sensitivity, see the text.}
    \label{fig:sensitivities-ALP}
\end{figure*}

The sensitivities to the benchmark models described in Sec.~\ref{sec:signal_char} are shown in Figures~\ref{fig:sensitivities-DP}-\ref{fig:sensitivities-ALP}. For comparison, the figures show the sensitivities of FASER and FASER2 experiments from~\cite{Antel:2023hkf}, as well as various LHCb searches from~\cite{LHCbCollaboration:2806113,LHCb:2018roe}.

The considered LLPs have very different phenomenology, which determines the different status of the exclusion of their parameter space by past experiments. For some of them, the unconstrained parameter space includes only the domain of large lifetimes $c\tau \gg 1\text{ m}$. For the other ones, lifetimes  $c\tau \lesssim 1\text{ m}$ also remain unexplored. It is because, on the one hand, limitations of the past prompt searches in luminosity and efficiency, which leaves small couplings unconstrained, and on the other hand, parametric smallness of the lifetime which prevented the past beam dump experiments with the far placement of the decay volume, e.g., CHARM, to be able to search for such LLPs. One of the powers of the \texttt{Downstream} setup is that it may search for LLPs in both these regimes.

For dark photons and $B-L$ mediators (Fig.~\ref{fig:sensitivities-DP}), the second scenario is realized. In particular, in the mass range $m_{V}\lesssim 0.6\text{ GeV}/c^2$, there is an underexplored parameter space of short lifetimes $c\tau\lesssim 1\text{ m}$. This mass range may be complementarily probed by various searches at LHCb, including the \texttt{Downstream} setup and the searches for resonance in di-electron and di-muon invariant mass restricted by VELO~\cite{LHCbCollaboration:2806113}. Depending on the luminosity, it may be able to search for masses $m_{V}\lesssim 1\text{ GeV}/c^2$. The upper bound of the sensitivities of FASER and FASER2 lies well below the sensitivity of \texttt{Downstream}, in good agreement with the estimate~\eqref{eq:upper-bound-faser}. The disconnected sensitivity regions in Fig.~\ref{fig:sensitivities-DP} appear due to the interplay between the behaviors of the LLP production rate and its lifetime. For these mediators, $c\tau \cdot g^{2}$ is parametrically very small, which requires a decrease in $g^{2}$ to make the LLPs possible to reach the decay volume before decaying. On the other hand, this would lead to a decrease in the production cross-section $\sigma_{pp\to\text{LLP}} \propto g^{2}$. Parametrically, the ratio $\sigma_{pp\to\text{LLP}}/g^{2}$ is too small in the mass range $0.5\text{ GeV}/c^2\lesssim m\lesssim 0.6\text{ GeV}/c^2$ to compensate for this decrease. However, it gets enhanced around the masses of $\rho/\omega$ mesons and their excitations (due to the mixing of the dark photons and $B$ mediator with $\omega,\rho,\phi$~\cite{Ilten:2018crw}).

Higgs-like scalars are efficiently produced by decays of $B$ mesons. Apart from using the \texttt{Downstream} setup, it may be possible to search for them at LHCb by studying processes of the type $B\to K^{(*)}+S(\to \mu\mu)$ localized in VELO, where $S$ would manifest itself via a resonant contribution in the dimuon invariant mass~\cite{LHCb:2015nkv,LHCb:2016awg}. Compared to the projections of the future reach of this type of search as reported in~\cite{LHCb:2018roe}, the \texttt{Downstream} setup would cover the lifetimes in two orders of magnitude larger (see Fig.~\ref{fig:sensitivities-Scalar}). The main reason for this is a suppression in the event rate by the reconstruction efficiency for $B\to K^{(*)}+S(\to \mu\mu)$ (coming from the $p_{T}$ cut on the outgoing muons, reconstruction of the kaon, and the requirement for the reconstructed $B$ decay vertex to be sufficiently displaced), the branching ratio $\text{Br}_{B\to K+S} \approx \text{Br}_{B\to X+S}/8$, the effective decay volume limited by VELO, and the branching ratio $S\to \mu\mu$ (remind Fig.~\ref{fig:decays}). 

As for the comparison with FASER/FASER2, for the model \textit{BC4} (zero trilinear coupling $hSS$), the obtained results are in agreement with the qualitative estimates made in Sec.~\ref{sec:comparison-lhc-based}. Compared to FASER, the \texttt{Downstream} setup may deliver a much better sensitivity. As for FASER2, the \texttt{Downstream} sensitivity would probe the same or slightly larger lifetimes at the lower bound, while for the upper bound, the probed domain is extended to the range of smaller lifetimes, thanks to a much shorter distance to the decay volume. In the case of a non-zero $hSS$ coupling (\textit{BC5}), scalars may be produced by the decays $B_{s}\to SS$ and $B\to SSX$ and the 2-body Higgs boson decays $h\to SS$. The experiment may be searching for such scalars up to the production threshold from Higgs bosons, $m_{S} < m_{h}/2$, again thanks to a very small distance to the decay volume. This is impossible at FASER, while the reach of FASER2 is limited to the vicinity of the kinematic threshold $m_{S}\simeq m_{h}/2$ due to the suppression in the number of scalars pointing to the detector~\cite{Boiarska:2019vid}.

For HNLs $N$ (Fig.~\ref{fig:sensitivities-HNL}), there are three mass domains depending on the main production channel -- by the decays of 
$D/\tau$ ($m_{N}\lesssim 2\text{ GeV}/c^2$), $B$ ($2\text{ GeV}/c^2\lesssim m_{N}\lesssim m_{B_{c}}-m_{l}$, where $l$ is the lepton corresponding to the HNL mixing), and $W$ ($m_{N}\gtrsim m_{B_{c}}$). The \texttt{Downstream} setup allows an efficient probe of the first two domains, with the maximal mass of the HNL being as large as $\simeq 20\text{ GeV}/c^2$. The HNLs produced by decays of $D/\tau$, following the kinematics of these particles, mainly point to the far-forward region not covered by LHCb. In comparison, FASER2 would be able to probe HNLs only up to masses $2\text{ GeV}/c^2\lesssim m_{N}\lesssim 3\text{ GeV}/c^2$, mainly because of its distant placement relative to the production point. 

Unlike the dark scalar case, there is no possibility to utilize the signature $B\to K + N(\to \mu\mu)$ for HNLs. First, HNLs are fermions, and the angular momentum conservation, together with the HNL interaction properties, requires the presence of an additional lepton in the $B$ decay. The probability of such process, $B_{s}\to K + N + \ell$, is very suppressed~\cite{Bondarenko:2018ptm}. Finally, the only HNL decay with the dimuon state is a three-body process $N\to \mu\mu\nu$; as a result, the dimuon mass distribution is not resonant.

The comparison with FASER/FASER2 shows the same pattern as in the case of dark scalars, again reproducing qualitative conclusions of Sec.~\ref{sec:comparison-lhc-based}.

For the ALPs with the universal coupling to fermions (Fig.~\ref{fig:sensitivities-ALP}), \textit{BC10}, the situation is very similar to the case of dark scalars since the dominant production channel is the same -- decays of $B$ mesons, while the decays into fermions have the similar Yukawa-like hierarchy: the corresponding decay width scales as $\Gamma_{a\to ff}\propto m_{f}^{2}$. The gaps in the sensitivity correspond to the vicinity of the masses of the neutral light mesons $m^{0} = \pi^{0},\eta,\eta'$ where the description of the ALP phenomenology based on the mixing with these mesons becomes inadequate. 

In the case of the ALPs coupled to gluons (\textit{BC11}), the mixing becomes the main production channel. This results in a worse sensitivity of the \texttt{Downstream} setup compared to FASER2. Indeed, $m^{0}$s have a very narrow angular distribution -- their characteristic $p_{T}$ is of the order of $\Lambda_{\text{QCD}}$. Given the typical energies of the order of TeV, the angular flux of mesons starts falling at $\theta<1\text{ mrad}$, i.e. well below the angular coverage of LHCb but within the range of FASER2. In addition, an important decay channel of these ALPs (in the mass range $m_{a}\lesssim m_{\eta}$) is into a pair of photons~\cite{Aloni:2018vki}, which are conservatively not considered as visible particles for the \texttt{Downstream} setup. Still, however, at the upper bound of the sensitivity, it would provide much better opportunities.

It is important to stress again (remind Sec.~\ref{sec:signal_char}) that the description of the ALP phenomenology considered in this paper differs from the description used to calculate the sensitivity of FASER2, which makes the direct comparison more complicated.

The sensitivities to all the LLPs considered in this paper may be improved if the effective decay volume extends from the end of the UT and until the SciFi layers. At present, work is being developed to include \textit{fareway} tracks, with only hits in the SciFi, and perform a fast vertexing at the HLT1, keeping a high throughput. This will extend the LLP search potential of LHCb even further.

\begin{figure*}[t!]
    \centering
    \includegraphics[width=0.8\textwidth]{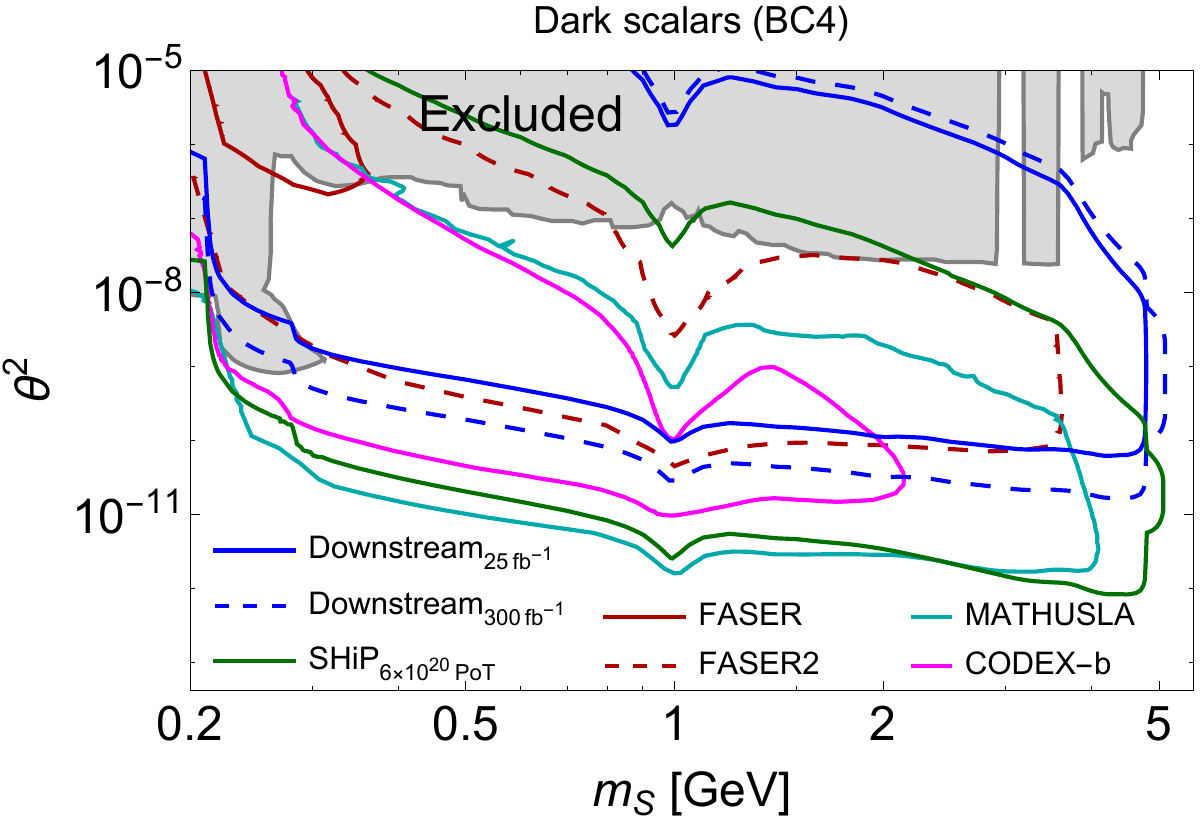}
    \caption{Comparison of the sensitivities of future proposed and approved experiments to the model of Higgs-like scalars (BC4). See text for details.}
    \label{fig:landscape}
\end{figure*}

Finally, it is important to consider the \texttt{Downstream} algorithm over a landscape of future experiments. As a reference model example for the comparison, Higgs-like scalars are chosen, because its production mode -- decays of $B$ mesons -- is representative for many other LLPs, such as HNLs and ALPs, and it may be possible to search for them at many experiments to be located at different facilities. The comparison of sensitivities is shown in Fig.~\ref{fig:landscape}. The included experiments are recently approved SHiP, FASER, FASER2, MATHUSLA, and CODEX-b. The sensitivity of CODEX-b is taken from~\cite{Antel:2023hkf}, the sensitivity of MATHUSLA from~\cite{Curtin:2023skh}, while the sensitivity of SHiP is computed using \texttt{SensCalc}. The comparison is tricky, since the experiments may fall into different categories: already approved or at the stage of proposals (CODEX-b, MATHUSLA, FASER2); be equipped with the full detector or with just tracking layers (MATHUSLA), which is crucial for identifying the LLP; to be running at different times. Namely, while the \texttt{Downstream} algorithm is going to be run already in 2024, while FASER is already collecting the data, the timescale for the other experiments is rather shifted: SHiP is expected to run after 2030~\cite{Aberle:2839677}, MATHUSLA and CODEX-b - during the High luminosity phase of the LHC~\cite{Antel:2023hkf}. This way, it is seen that the \texttt{Downstream} algorithm is the best experiment to search for LLPs in the next few years.

\section{Conclusions}
\label{sec:conclusions}

The current search strategies employed at the LHC's primary detectors, namely ATLAS, CMS, and LHCb, are not well-suited for exploring the parameter space associated with hypothetical long-lived particles (LLPs) in the GeV mass range. Consequently, there has been a surge in proposals for experiments beyond the LHC dedicated to the search for LLPs. This study demonstrates the potential of efficiently harnessing the capabilities of the LHCb experiment by implementing a novel \texttt{Downstream} algorithm. This approach enables the exploration of events lacking hits in the innermost LHCb tracker. In comparison to the existing search methods employed by LHCb, this algorithm offers the advantages of triggering at the production vertex, enhanced background control, an expanded effective decay volume, and the ability to investigate various final states resulting from the decays of LLPs.

The \texttt{Downstream} setup holds promise for the investigation of a diverse range of LLPs, potentially rivaling the exploration potential of established LHC-based experiments like FASER2 (refer to Sec.~\ref{sec:signal_char}). Leveraging the complete dataset from LHCb until Run 6, it becomes feasible to probe Heavy Neutral Leptons (HNLs) with masses up to approximately $20\text{ GeV}/c^2$, as well as dark photons and $B-L$ mediators with masses of around $1\text{ GeV}/c^2$. Moreover, this approach extends the search to Higgs-like scalars with lifetimes exceeding those accessible by the current LHCb search strategies, and to axion-like particles with various coupling patterns (as outlined in Sec.~\ref{sec:sensitivity}). Further enhancements in sensitivity can be achieved by enlarging the effective decay volume and incorporating the possibility of reconstructing final states comprising exclusively photons, contingent upon the development of new triggers.

\section*{Acknowledgements}
The authors thank Andrii Usachov for the useful discussions concerning the experimental part. Part of this work is supported by the State Agency of Research, Spanish Ministry of Innovation and Research and European Union-NextGenEU. The authors acknowledge the support from project TED2021-130852B-I00 and CONEXION AIHUB-CSIC, and from the European Union's Horizon 2020 research and innovation program under the Marie Sklodowska-Curie grant agreement No. 860881-HIDDeN. Support for V.S. and V.K. was provided by the US NSF cooperative agreement OAC-1836650 (IRIS-HEP) and the Simons Foundation.
\clearpage

\appendix

\section{Implementation of the setup for the \texttt{Downstream} algorithm in \texttt{SensCalc}}
\label{app:senscalc}

\begin{figure*}[t!]
    \centering
    \includegraphics[width=0.7\textwidth]{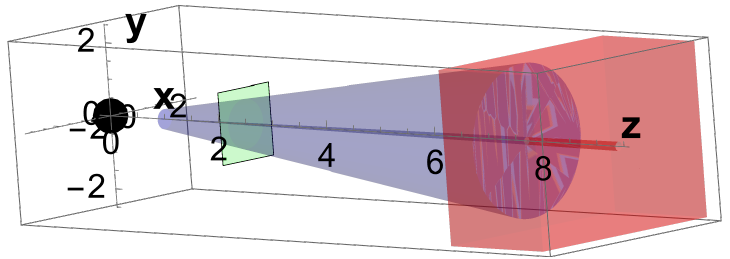}
    \caption{The geometry of the LHCb as implemented in \texttt{SensCalc}. The thick black point corresponds to the origin of the coordinate frame, coinciding with the point of $pp$ collisions. The blue region corresponds to the decay volume, while the red one is the detector. The green plane shows the location of the UT layers; if the tracks are also required to intersect the UT, the decay volume shrinks to the domain until the UT plane.}
    \label{fig:setup-senscalc}
\end{figure*}

The LHCb with the \texttt{Downstream} setup has been implemented in the \texttt{SensCalc} framework to estimate the number of events and allow for comparisons. The implementation is shown in Fig.~\ref{fig:setup-senscalc}, and details are given in the following. 

For the decay volume, conical frustum covering pseudorapidities $2<\eta<5$ and located in the longitudinal displacement $z$ from $z_{\text{min}} = 1\text{ m}$ to $z_{\text{max}} = 7.7\text{ m}$ is considered, where the first SciFi layer is located. If the tracks must also intersect the UT, the size of the decay volume shrinks to $z_{\text{max}}\approx 2.5\text{ m}$, which is the beginning of the UT. For the geometry of the Sci-Fi layers, a parallelepiped with dimensions $6.48\text{ m}\times 4.83\text{ m}\times 1.7\text{ m}$ with a hole of the radius $R = 9\text{ cm}$ to account for the beam pipe is used, following~\cite{LHCbU1}. The magnetic field of the dipole magnet is extended from $z = 3.5\text{ m}$ to $z = 7.5\text{ m}$, with the integrated field $\int B dl = 4\text{ T $\cdot$ m}$.

The setup is available with the current \texttt{SensCalc} repository~\cite{maksym_ovchynnikov_2023_10001210}. Depending on details, there are three implemented options: 
\begin{itemize}
    \item[--] \texttt{LHCb-downstream} 
    \item[--] \texttt{LHCb-downstream-T-tracks-only}
    \item[--] \texttt{LHCb-downstream-full}
\end{itemize}
The first one corresponds to the setup considered in this paper -- the one with the decay volume extending from $z = 1\text{ m}$ to $z = 2.5\text{ m}$ and SciFi as the detector. The second option also includes the domain $2.5<z<z_{\text{SciFi}}$ as the decay volume; it corresponds to the scenario when the event may be reconstructed purely by T tracks. Finally, the last one is a sketch of the full LHCb detector up to muon stations (see Fig.~\ref{fig:lhcb}). Users may easily add new configurations or modify the existing ones.

\subsection{Validation}
\label{app:senscalc-validation}
To validate the prediction of \texttt{SensCalc}, the event rate for the dark scalar mixed with the standard Higgs boson is analysed. Specifically, the acceptance for the dark scalars to have $2<\eta<5$ and the $z$-dependence of the pure geometric part of the decay products acceptance (i.e., with $\epsilon_{\text{rec}} = 1$) is studied, which is defined as

\begin{equation}
            \epsilon_{\text{det}}(m_{S},z_{S}) \equiv \frac{\langle f_{\text{LLP}}\frac{dP_{\text{decay}}}{dz}\epsilon_{\text{decay}}\rangle_{\theta,E}}{\langle f_{\text{LLP}}\frac{dP_{\text{decay}}}{dz}\rangle_{\theta,E}},
            \label{eq:decay-acceptance}
\end{equation}
\noindent and results compared with the LHCb simulations. 

Simulations in this work are performed using a specific package called \texttt{RapidSim} \cite{Cowan_2017}, an application for fast simulation of phase space decays of heavy hadrons, which allows for quick studies of the properties of signal and background decays in particle physics analyses. It includes realistic production kinematic distributions, efficiencies, and momentum resolutions.

\begin{figure}[!h]
    \centering
    \includegraphics[width=0.45\textwidth]{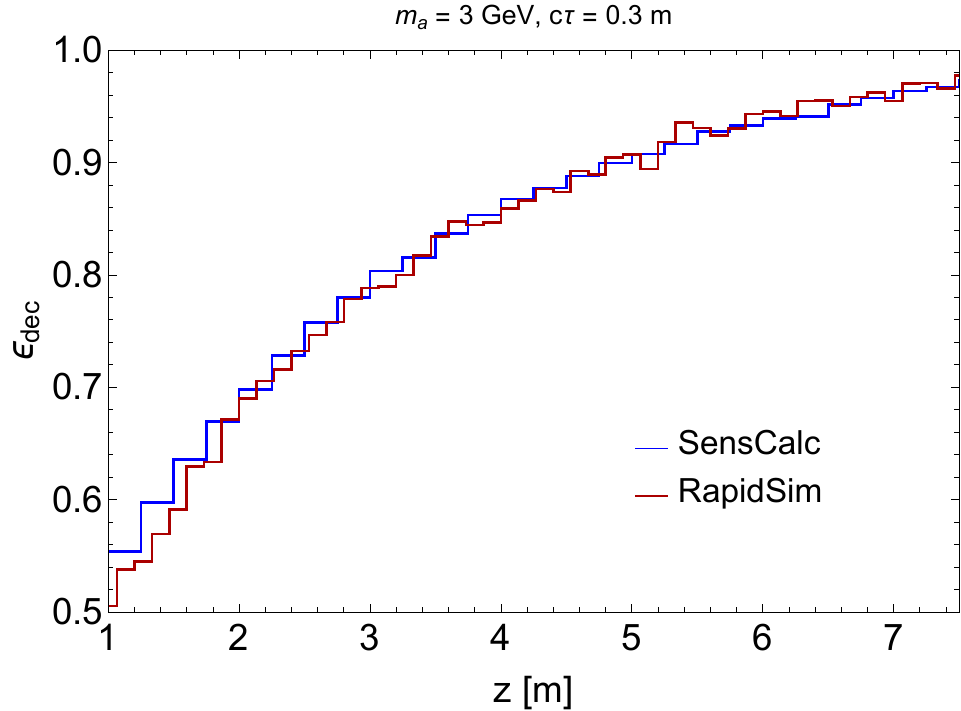}
    \caption{The behavior of the decay products acceptance~\eqref{eq:decay-acceptance} if assuming $\epsilon_{\text{rec}} = 1$, as estimated by \texttt{SensCalc} (blue) and predicted by the RapidSim simulations (red) \cite{Cowan_2017}. See text for details.}
    \label{fig:validation}
\end{figure}

As it is shown in Fig.~\ref{fig:validation}, a good agreement is obtained between the acceptance predicted by \texttt{SensCalc} and the \texttt{RapidSim} simulation.

\clearpage

\bibliographystyle{utphys} 
\bibliography{bib}

\providecommand{\href}[2]{#2}\begingroup\raggedright\begin{thebibliography}{10}

\bibitem{Alekhin:2015byh}
S.~Alekhin {\em et~al.}, ``{A facility to Search for Hidden Particles at the
  CERN SPS: the SHiP physics case},''
  \href{https://dx.doi.org/10.1088/0034-4885/79/12/124201}{{\em Rept. Prog.
  Phys.} {\bfseries 79} no.~12, (2016) 124201},
  \href{https://arxiv.org/abs/1504.04855}{{\ttfamily arXiv:1504.04855
  [hep-ph]}}.

\bibitem{Beacham:2019nyx}
J.~Beacham {\em et~al.}, ``{Physics Beyond Colliders at CERN: Beyond the
  Standard Model Working Group Report},''
  \href{https://dx.doi.org/10.1088/1361-6471/ab4cd2}{{\em J. Phys. G}
  {\bfseries 47} no.~1, (2020) 010501},
  \href{https://arxiv.org/abs/1901.09966}{{\ttfamily arXiv:1901.09966
  [hep-ex]}}.

\bibitem{Antel:2023hkf}
C.~Antel {\em et~al.}, ``{Feebly Interacting Particles: FIPs 2022 workshop
  report},'' in {\em {Workshop on Feebly-Interacting Particles}}.
\newblock 5, 2023.
\newblock \href{https://arxiv.org/abs/2305.01715}{{\ttfamily arXiv:2305.01715
  [hep-ph]}}.

\bibitem{CMS:2022fut}
{\bfseries CMS} Collaboration, A.~Tumasyan {\em et~al.}, ``{Search for
  long-lived heavy neutral leptons with displaced vertices in proton-proton
  collisions at $ \sqrt{\mathrm{s}} $ =13 TeV},''
  \href{https://dx.doi.org/10.1007/JHEP07(2022)081}{{\em JHEP} {\bfseries 07}
  (2022) 081}, \href{https://arxiv.org/abs/2201.05578}{{\ttfamily
  arXiv:2201.05578 [hep-ex]}}.

\bibitem{ATLAS:2022atq}
{\bfseries ATLAS} Collaboration, G.~Aad {\em et~al.}, ``{Search for Heavy
  Neutral Leptons in Decays of W Bosons Using a Dilepton Displaced Vertex in
  s=13\,\,TeV pp Collisions with the ATLAS Detector},''
  \href{https://dx.doi.org/10.1103/PhysRevLett.131.061803}{{\em Phys. Rev.
  Lett.} {\bfseries 131} no.~6, (2023) 061803},
  \href{https://arxiv.org/abs/2204.11988}{{\ttfamily arXiv:2204.11988
  [hep-ex]}}.

\bibitem{LHCb:impact}
L.Calefice, A.Hennequin, L.Henry, B.Jashal, D.Mendoza, A.Oyanguren,
  I.Sanderswood, C.Vázquez, and J.Zhuo, ``{Effect of the high-level trigger
  for detecting long-lived particles at LHCb},'' {\em {Front. Big Data, Sec.
  Big Data and AI in High Energy Physics}} {\bfseries 5} (2022) .
  \url{https://www.frontiersin.org/articles/10.3389/fdata.2022.1008737/full}.

\bibitem{Shchutska:2023cgi}
L.~Shchutska and M.~Ovchynnikov, ``{Prospects for LLP searches at the LHC in
  Run 3 and HL-LHC},'' \href{https://dx.doi.org/10.22323/1.422.0094}{{\em PoS}
  {\bfseries LHCP2022} (2023) 094}.

\bibitem{CMS-PAS-EXO-22-017}
{\bfseries CMS} Collaboration, ``{Search for long-lived heavy neutral leptons
  decaying in the CMS muon detectors in proton-proton collisions at
  $\sqrt{s}=13~\mathrm{TeV}$},'' tech. rep., CERN, Geneva, 2023.
\newblock \url{https://cds.cern.ch/record/2865227}.

\bibitem{CERN-LHCC-2020-006}
{\bfseries LHCb} Collaboration,
  \href{https://dx.doi.org/10.17181/CERN.QDVA.5PIR}{``{LHCb Upgrade GPU High
  Level Trigger Technical Design Report},''} tech. rep., CERN, Geneva, 2020.
\newblock \url{https://cds.cern.ch/record/2717938}.

\bibitem{mrdownstream}
{Jashal~B., Oyanguren~A., Zhuo~J.}, ``Downstream/longlivedparticle track
  reconstruction.'' 2023.
\newblock Available online at:
  \url{https://gitlab.cern.ch/lhcb/Allen/-/merge_requests/1095}.

\bibitem{LHCbU1}
{\bfseries LHCb} Collaboration, I.~Bediaga {\em et~al.}, ``{Framework TDR for
  the LHCb Upgrade: Technical Design Report},'' tech. rep., 2012.
\newblock \url{https://cds.cern.ch/record/1443882}.

\bibitem{LHCb_2008}
T.~L. Collaboration, A.~A.~A. Jr, {\em et~al.}, ``The lhcb detector at the
  lhc,'' \href{https://dx.doi.org/10.1088/1748-0221/3/08/S08005}{{\em Journal
  of Instrumentation} {\bfseries 3} no.~08, (Aug, 2008) S08005}.
  \url{https://dx.doi.org/10.1088/1748-0221/3/08/S08005}.

\bibitem{LHCb:2023hlw}
{\bfseries LHCb} Collaboration, R.~Aaij {\em et~al.}, ``{The LHCb upgrade I},''
  \href{https://arxiv.org/abs/2305.10515}{{\ttfamily arXiv:2305.10515
  [hep-ex]}}.

\bibitem{arXiv:2211.10920}
{\bfseries LHCb} Collaboration, ``{Long-lived particle reconstruction
  downstream of the LHCb magnet},'' tech. rep., 2022.
\newblock \href{https://arxiv.org/abs/2211.10920}{{\ttfamily
  arXiv:2211.10920}}.
\newblock \url{https://cds.cern.ch/record/2841793}.
\newblock All figures and tables, along with machine-readable versions and any
  supplementary material and additional information, are available at
  https://cern.ch/lhcbproject/Publications/p/LHCb-DP-2022-001.html (LHCb public
  pages).

\bibitem{Li:2752971}
P.~Li, E.~Rodrigues, and S.~Stahl, ``{Tracking Definitions and Conventions for
  Run 3 and Beyond},'' tech. rep., CERN, Geneva, 2021.
\newblock \url{https://cds.cern.ch/record/2752971}.

\bibitem{CERN-LHCC-2014-016}
``{LHCb Trigger and Online Upgrade Technical Design Report},'' tech. rep.,
  2014.
\newblock \url{https://cds.cern.ch/record/1701361}.

\bibitem{Aaij:2019zbu}
R.~Aaij {\em et~al.}, ``{Allen: A high level trigger on GPUs for LHCb},''
  \href{https://dx.doi.org/10.1007/s41781-020-00039-7}{{\em Comput. Softw. Big
  Sci.} {\bfseries 4} no.~1, (2020) 7},
  \href{https://arxiv.org/abs/1912.09161}{{\ttfamily arXiv:1912.09161
  [physics.ins-det]}}.

\bibitem{Jashal:2881886}
B.~K. Jashal, ``{Triggering new discoveries: development of advanced HLT1
  algorithms for detection of long-lived particles at LHCb}.'' 2023.
\newblock \url{https://cds.cern.ch/record/2881886}. Presented 07 Nov 2023.

\bibitem{LHCB-FIGURE-2023-028}
{\bfseries LHCb} Collaboration, ``{Downstream track reconstruction in HLT1},''.
  \url{https://cds.cern.ch/record/2875269}.

\bibitem{SHiP:2020vbd}
{\bfseries SHiP} Collaboration, C.~Ahdida {\em et~al.}, ``{Sensitivity of the
  SHiP experiment to dark photons decaying to a pair of charged particles},''
  \href{https://dx.doi.org/10.1140/epjc/s10052-021-09224-3}{{\em Eur. Phys. J.
  C} {\bfseries 81} no.~5, (2021) 451},
  \href{https://arxiv.org/abs/2011.05115}{{\ttfamily arXiv:2011.05115
  [hep-ex]}}.

\bibitem{Ilten:2018crw}
P.~Ilten, Y.~Soreq, M.~Williams, and W.~Xue, ``{Serendipity in dark photon
  searches},'' \href{https://dx.doi.org/10.1007/JHEP06(2018)004}{{\em JHEP}
  {\bfseries 06} (2018) 004},
  \href{https://arxiv.org/abs/1801.04847}{{\ttfamily arXiv:1801.04847
  [hep-ph]}}.

\bibitem{Boiarska:2019jym}
I.~Boiarska, K.~Bondarenko, A.~Boyarsky, V.~Gorkavenko, M.~Ovchynnikov, and
  A.~Sokolenko, ``{Phenomenology of GeV-scale scalar portal},''
  \href{https://dx.doi.org/10.1007/JHEP11(2019)162}{{\em JHEP} {\bfseries 11}
  (2019) 162}, \href{https://arxiv.org/abs/1904.10447}{{\ttfamily
  arXiv:1904.10447 [hep-ph]}}.

\bibitem{Bondarenko:2018ptm}
K.~Bondarenko, A.~Boyarsky, D.~Gorbunov, and O.~Ruchayskiy, ``{Phenomenology of
  GeV-scale Heavy Neutral Leptons},''
  \href{https://dx.doi.org/10.1007/JHEP11(2018)032}{{\em JHEP} {\bfseries 11}
  (2018) 032}, \href{https://arxiv.org/abs/1805.08567}{{\ttfamily
  arXiv:1805.08567 [hep-ph]}}.

\bibitem{DallaValleGarcia:2023xhh}
G.~Dalla Valle~Garcia, F.~Kahlhoefer, M.~Ovchynnikov, and A.~Zaporozhchenko,
  ``{Phenomenology of axion-like particles with universal fermion couplings --
  revisited},'' \href{https://arxiv.org/abs/2310.03524}{{\ttfamily
  arXiv:2310.03524 [hep-ph]}}.

\bibitem{Aloni:2018vki}
D.~Aloni, Y.~Soreq, and M.~Williams, ``{Coupling QCD-Scale Axionlike Particles
  to Gluons},'' \href{https://dx.doi.org/10.1103/PhysRevLett.123.031803}{{\em
  Phys. Rev. Lett.} {\bfseries 123} no.~3, (2019) 031803},
  \href{https://arxiv.org/abs/1811.03474}{{\ttfamily arXiv:1811.03474
  [hep-ph]}}.

\bibitem{Ovchynnikov:2023cry}
M.~Ovchynnikov, J.-L. Tastet, O.~Mikulenko, and K.~Bondarenko, ``{Sensitivities
  to feebly interacting particles: public and unified calculations},''
  \href{https://dx.doi.org/10.1103/PhysRevD.108.075028}{{\em Phys. Rev. D}
  {\bfseries 108} no.~7, (5, 2023) 075028},
  \href{https://arxiv.org/abs/2305.13383}{{\ttfamily arXiv:2305.13383
  [hep-ph]}}.

\bibitem{Mikulenko:2023}
O.~Mikulenko, K.~Bondarenko, A.~Boyarsky, and O.~Ruchayskiy, ``{Unveiling new
  physics with discoveries at Intensity Frontier},''
  \href{https://arxiv.org/abs/to appear}{{\ttfamily arXiv:to appear [hep-ph]}}.

\bibitem{LHCb:2016awg}
{\bfseries LHCb} Collaboration, R.~Aaij {\em et~al.}, ``{Search for long-lived
  scalar particles in $B^+ \to K^+ \chi (\mu^+\mu^-)$ decays},''
  \href{https://dx.doi.org/10.1103/PhysRevD.95.071101}{{\em Phys. Rev. D}
  {\bfseries 95} no.~7, (2017) 071101},
  \href{https://arxiv.org/abs/1612.07818}{{\ttfamily arXiv:1612.07818
  [hep-ex]}}.

\bibitem{Mazurek:2022tlu}
M.~Mazurek, M.~Clemencic, and G.~Corti, ``{Gauss and Gaussino: the LHCb
  simulation software and its new experiment agnostic core framework},''
  \href{https://dx.doi.org/10.22323/1.414.0225}{{\em PoS} {\bfseries ICHEP2022}
  (11, 2022) 225}.

\bibitem{Alexander:2018png}
M.~Alexander {\em et~al.}, ``{Mapping the material in the LHCb vertex locator
  using secondary hadronic interactions},''
  \href{https://dx.doi.org/10.1088/1748-0221/13/06/P06008}{{\em JINST}
  {\bfseries 13} no.~06, (2018) P06008},
  \href{https://arxiv.org/abs/1803.07466}{{\ttfamily arXiv:1803.07466
  [physics.ins-det]}}.

\bibitem{Bondarenko:2019yob}
K.~Bondarenko, A.~Boyarsky, M.~Ovchynnikov, and O.~Ruchayskiy, ``{Sensitivity
  of the intensity frontier experiments for neutrino and scalar portals:
  analytic estimates},'' \href{https://dx.doi.org/10.1007/JHEP08(2019)061}{{\em
  JHEP} {\bfseries 08} (2019) 061},
  \href{https://arxiv.org/abs/1902.06240}{{\ttfamily arXiv:1902.06240
  [hep-ph]}}.

\bibitem{FASER:2018bac}
{\bfseries FASER} Collaboration, A.~Ariga {\em et~al.}, ``{Technical Proposal
  for FASER: ForwArd Search ExpeRiment at the LHC},''
  \href{https://arxiv.org/abs/1812.09139}{{\ttfamily arXiv:1812.09139
  [physics.ins-det]}}.

\bibitem{FASER:2018ceo}
{\bfseries FASER} Collaboration, A.~Ariga {\em et~al.}, ``{Letter of Intent for
  FASER: ForwArd Search ExpeRiment at the LHC},''
  \href{https://arxiv.org/abs/1811.10243}{{\ttfamily arXiv:1811.10243
  [physics.ins-det]}}.

\bibitem{Feng:2022inv}
J.~L. Feng {\em et~al.}, ``{The Forward Physics Facility at the High-Luminosity
  LHC},'' \href{https://dx.doi.org/10.1088/1361-6471/ac865e}{{\em J. Phys. G}
  {\bfseries 50} no.~3, (2023) 030501},
  \href{https://arxiv.org/abs/2203.05090}{{\ttfamily arXiv:2203.05090
  [hep-ex]}}.

\bibitem{Feldman:1997qc}
G.~J. Feldman and R.~D. Cousins, ``{A Unified approach to the classical
  statistical analysis of small signals},''
  \href{https://dx.doi.org/10.1103/PhysRevD.57.3873}{{\em Phys. Rev. D}
  {\bfseries 57} (1998) 3873--3889},
  \href{https://arxiv.org/abs/physics/9711021}{{\ttfamily
  arXiv:physics/9711021}}.

\bibitem{Bityukov:2000ms}
S.~I. Bityukov and N.~V. Krasnikov, ``{Confidence intervals for the parameter
  of Poisson distribution in presence of background},''
  \href{https://arxiv.org/abs/physics/0009064}{{\ttfamily
  arXiv:physics/0009064}}.

\bibitem{LHCbCollaboration:2806113}
L.~E. LHCb~Collaboration, ``{Future physics potential of LHCb},'' tech. rep.,
  CERN, Geneva, 2022.
\newblock \url{https://cds.cern.ch/record/2806113}.

\bibitem{LHCb:2018roe}
{\bfseries LHCb} Collaboration, R.~Aaij {\em et~al.}, ``{Physics case for an
  LHCb Upgrade II - Opportunities in flavour physics, and beyond, in the HL-LHC
  era},'' \href{https://arxiv.org/abs/1808.08865}{{\ttfamily arXiv:1808.08865
  [hep-ex]}}.

\bibitem{LHCb:2015nkv}
{\bfseries LHCb} Collaboration, R.~Aaij {\em et~al.}, ``{Search for
  hidden-sector bosons in $B^0 \!\to K^{*0}\mu^+\mu^-$ decays},''
  \href{https://dx.doi.org/10.1103/PhysRevLett.115.161802}{{\em Phys. Rev.
  Lett.} {\bfseries 115} no.~16, (2015) 161802},
  \href{https://arxiv.org/abs/1508.04094}{{\ttfamily arXiv:1508.04094
  [hep-ex]}}.

\bibitem{Boiarska:2019vid}
I.~Boiarska, K.~Bondarenko, A.~Boyarsky, M.~Ovchynnikov, O.~Ruchayskiy, and
  A.~Sokolenko, ``{Light scalar production from Higgs bosons and FASER 2},''
  \href{https://dx.doi.org/10.1007/JHEP05(2020)049}{{\em JHEP} {\bfseries 05}
  (2020) 049}, \href{https://arxiv.org/abs/1908.04635}{{\ttfamily
  arXiv:1908.04635 [hep-ph]}}.

\bibitem{Boyarsky:2020dzc}
A.~Boyarsky, M.~Ovchynnikov, O.~Ruchayskiy, and V.~Syvolap, ``{Improved big
  bang nucleosynthesis constraints on heavy neutral leptons},''
  \href{https://dx.doi.org/10.1103/PhysRevD.104.023517}{{\em Phys. Rev. D}
  {\bfseries 104} no.~2, (2021) 023517},
  \href{https://arxiv.org/abs/2008.00749}{{\ttfamily arXiv:2008.00749
  [hep-ph]}}.

\bibitem{Sabti:2020yrt}
N.~Sabti, A.~Magalich, and A.~Filimonova, ``{An Extended Analysis of Heavy
  Neutral Leptons during Big Bang Nucleosynthesis},''
  \href{https://dx.doi.org/10.1088/1475-7516/2020/11/056}{{\em JCAP} {\bfseries
  11} (2020) 056}, \href{https://arxiv.org/abs/2006.07387}{{\ttfamily
  arXiv:2006.07387 [hep-ph]}}.

\bibitem{Curtin:2023skh}
D.~Curtin and J.~S. Grewal, ``{Long Lived Particle Decays in MATHUSLA},''
  \href{https://arxiv.org/abs/2308.05860}{{\ttfamily arXiv:2308.05860
  [hep-ph]}}.

\bibitem{Aberle:2839677}
{\bfseries SHiP} Collaboration, O.~e.~a. Aberle, ``{BDF/SHiP at the ECN3
  high-intensity beam facility},'' tech. rep., CERN, Geneva, 2022.
\newblock \url{https://cds.cern.ch/record/2839677}.

\bibitem{maksym_ovchynnikov_2023_10001210}
M.~Ovchynnikov, ``Senscalc.'' Oct., 2023.

\bibitem{Cowan_2017}
G.~Cowan, D.~Craik, and M.~Needham, ``Rapidsim: An application for the fast
  simulation of heavy-quark hadron decays,''
  \href{https://dx.doi.org/10.1016/j.cpc.2017.01.029}{{\em Computer Physics
  Communications} {\bfseries 214} (May, 2017) 239–246}.
  \url{http://dx.doi.org/10.1016/j.cpc.2017.01.029}.

\end{thebibliography}\endgroup

\end{document}